\documentclass[11pt]{article}
\usepackage[margin=1in]{geometry}
\usepackage{amsmath,amssymb}
\usepackage{graphicx}
\usepackage{booktabs}
\usepackage{multirow}
\usepackage{xcolor}
\usepackage{cite}
\usepackage{url}
\usepackage[hidelinks]{hyperref}

\newcommand{\prov}[1]{#1}
\newcommand{\lam}{\lambda}
\newcommand{\rrb}{r_{\mathrm{rb}}}
\newcommand{\figfile}[2]{%
  \IfFileExists{figures/#1.pdf}%
    {\includegraphics[width=#2]{figures/#1.pdf}}%
    {\fbox{\parbox[c][4cm][c]{\dimexpr#2-2\fboxsep-2\fboxrule\relax}{\centering\small Placeholder: \texttt{\detokenize{figures/#1.pdf}}}}}}
\title{Koopman Operator Analysis of EEG Microstate Dynamics Reveals a Cross-Dataset, Hearing-Level Adjusted Signature of Tinnitus}
\author{\parbox{\dimexpr\textwidth-2\tabcolsep\relax}{\centering Adyant~Balaji$^{1}$, Abhinav~Uppal$^{1}$, Min~Suk~Lee$^{1}$, Yuchen~Xu$^{1}$, Akihiro~J. Matsuoka$^{2}$, Gert~Cauwenberghs$^{1}$}}
\date{}

\begin{document}
\maketitle
\begin{center}\small
$^{1}$Department of Bioengineering, University of California San Diego, La Jolla, CA 92093 USA\\
$^{2}$Department of Otolaryngology and Head and Neck Surgery, University of California San Diego School of Medicine, San Diego, CA 92037, USA\\
Correspondence: abalaji@ucsd.edu
\end{center}
\linespread{1.6}\selectfont

\begin{abstract}
While major progress has been made in identifying EEG biomarkers for chronic Tinnitus, hearing loss remains as a major confound that must be accounted for in the identified effect. These markers have rarely been tested across independent datasets, a further test of robustness. We analyzed resting-state EEG microstate statistics, and developed a Koopman-operator description of the same microstate signal, carry a tinnitus-related difference that replicates across sites and survives adjustment for hearing level. Three public resting-state datasets (335 subjects; 234 tinnitus, 101 controls) were projected onto a fixed seven-map microstate template. Pooled across datasets, \prov{2/56} microstate statistics differed consistently, and after adjustment for pure-tone average (PTA) \prov{none} remained significant after FDR correction. Tinnitus subjects had a larger fraction of fast-decaying Koopman modes in all three datasets, and this difference survived PTA adjustment (pooled $r=$ \prov{0.26}, $p=$ \prov{0.002}). Microstate differences in tinnitus are substantially explained by hearing level; operator-level dynamics of the same signal retain a hearing-adjusted difference that generalizes across sites.
\end{abstract}

\section*{Introduction}
Tinnitus is the persistent perception of sound in the absence of an external acoustic stimulus. It affects approximately 10\% of US adults, with a significant proportion reporting severe distress, sleep disturbance, and cognitive impairment \cite{bhatt_prevalence_2016, tunkel_clinical_2014}. Despite its prevalence and impact on quality of life, tinnitus lacks a validated objective biomarker. Diagnosis remains entirely subjective, treatment outcomes are inconsistent, and patient stratification for clinical trials is limited. This motivates the search for reliable, generalizable neural signatures that can be measured non-invasively.

A substantial obstacle to this goal is hearing loss, which commonly accompanies tinnitus. In a population cohort, hearing impairment more than doubled the odds of tinnitus (OR 2.27) \cite{oosterloo_prevalence_2021}. The central gain model \cite{norena_integrative_2011} proposes that peripheral hearing loss triggers maladaptive upregulation of central neural activity, resulting in the phantom percept. Accordingly, tinnitus cohorts typically have poorer hearing than their controls, and hearing loss itself alters resting brain activity \cite{chen_resting-state_2026, cai_altered_2019}. A neural marker that is not controlled for hearing level may therefore index hearing loss rather than tinnitus. 

Resting-state electroencephalography (EEG) has emerged as a leading modality for investigating tinnitus-related neural correlates, owing to its high temporal resolution, low cost, and clinical accessibility. Converging evidence from EEG and magnetoencephalography studies identifies several recurrent findings. Enhanced delta and gamma and reduced alpha power in auditory cortex \cite{weisz_tinnitus_2005, loo_tinnitus_2009}, and distributed connectivity changes linking auditory, prefrontal, and limbic regions \cite{de_ridder_phantom_2011, elgoyhen_tinnitus_2015}. Nevertheless, translating these group-level findings into individual-level biomarkers remains challenging, largely because the extracted features tend to be sensitive to recording hardware, protocol, and population, and hearing level is not consistently controlled. These effects have also never been compared across datasets. 

EEG microstate analysis characterizes the quasi-stable topographic configurations through which the scalp field evolves on the millisecond timescale \cite{lehmann_eeg_1987, michel_eeg_2018}. Each microstate represents a period of near-stationary scalp potential, and the sequence of states (their coverage, durations and transition probabilities) encodes the temporal structure of spontaneous large-scale network dynamics. Altered microstate dynamics have been reported in idiopathic sudden sensorineural hearing loss and in tinnitus \cite{cai_altered_2019, raeisi_enhanced_2025}, suggesting potential biomarkers, but these do not separate tinnitus from hearing loss.

Beyond spectral and microstate approaches, the Koopman operator framework represents nonlinear dynamics in a globally linear space \cite{koopman_hamiltonian_1931, mezic_spectral_2005, budisic_applied_2012}. Observable functions of the system state evolve under an infinite-dimensional linear operator. Dynamic mode decomposition (DMD) and its extended and Hankel (time-delay) variants \cite{schmid_dynamic_2010, williams_data_2015, arbabi_ergodic_2017, wanner_robust_2022} give finite-rank, data-driven approximations of this operator, whose complex eigenvalues encode the decay rate and frequency of the dynamics. Koopman and DMD methods have recently been applied to neural data \cite{brunton_extracting_2016, marrouch_data-driven_2020, qian_koopman-operator-theoretical_2021}. Applied to the microstate activations themselves, they describe how the scalp field moves within and between states, at the resolution of single samples rather than discrete state segments.

In this work we systematically evaluate feature extraction frameworks applied to resting-state EEG from three independently collected datasets totaling 234 tinnitus and 101 control subjects, with channel-level EEG spectral power as a baseline reference. Here, we show that tinnitus and control subjects largely share microstate topographies, and that the microstate statistics that differ between them, pooled across datasets with a fixed microstate template, are partly explained by hearing level and do not replicate after adjusting for it. In contrast, a Koopman-operator representation of the same microstate signal reveals an eigenvalue spectrum with a consistent shift toward faster-decaying modes in tinnitus that persists after adjustment for hearing level. A classifier built on these eigenvalues also generalized across datasets better than a classifier using microstate statistics. 

% =====================================================================

\section*{Results}
\label{sec:results}

\begin{figure*}[!htbp]
\centering
% \figfile{fig1_overview}{\textwidth}
\includegraphics[width=\textwidth]{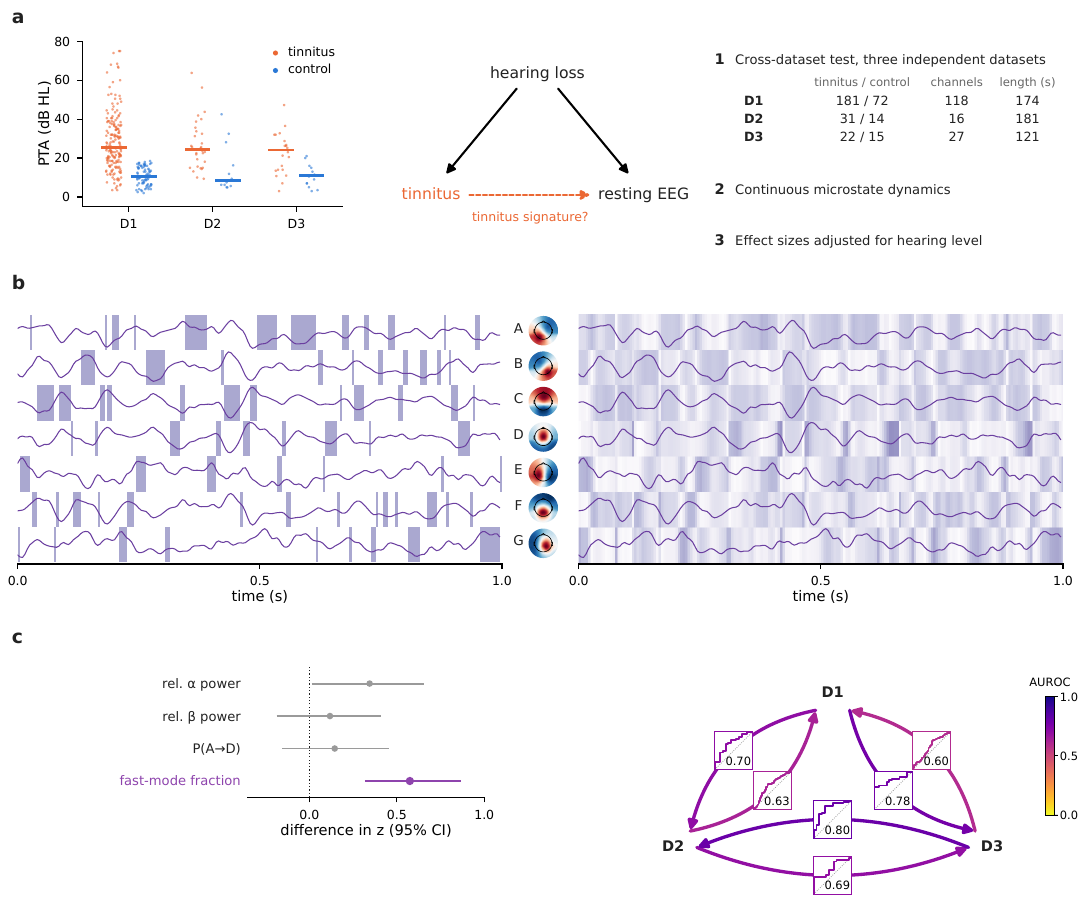}
\caption{\scriptsize Study overview. (a) Tinnitus comes with hearing loss. Left: pure-tone average (PTA) of each subject by group and dataset (bars: medians); PTA alone separates the groups with AUROC 0.78--0.90 (Table~\ref{tab:hearing}). Middle: hearing loss is associated both with tinnitus and with changes in the resting EEG, so an EEG group difference may reflect hearing level rather than tinnitus. Right: Study contributions (b) Two views of the same microstate activations $\mathbf{a}(t)$, shown for 1~s of a D2 tinnitus subject of Fig.~\ref{fig:pipeline}. The discrete view labels each sample by its most active map and summarizes the sequence by coverage, duration and transition probabilities (56 statistics). The continuous view fits a Koopman operator to the quadratic, delay-embedded activations; (c) Left: tinnitus minus control at the same hearing level. Each feature was z-scored within its dataset, the group difference was taken within each hearing-matched PTA bin ($\leq10$, 10--15 and 15--20~dB; \prov{74} tinnitus, \prov{95} control subjects) and averaged over the bins; bars: 95\% bootstrap CIs, resampling subjects within bin and group. Right: cross-dataset classification with the Koopman features. Each curved arrow is one train$\to$test direction, colored by its AUROC, with its ROC curve as an inset at its midpoint (Table~\ref{tab:clf}).}
\label{fig:overview}
\end{figure*}

% \begin{figure}[!htbp]
% \centering
% \figfile{fig2_custo_template}{\columnwidth}
% \caption{The Custo et al.\ $k=7$ microstate template (maps A--G) on its 204-channel montage. The same maps, spline-interpolated onto each dataset's montage, were used for every analysis.}
% \label{fig:maps}
% \end{figure}

\begin{figure*}[!htbp]
\centering
% \figfile{fig1_koopman_pipeline}{\textwidth}
\includegraphics[width=\textwidth]{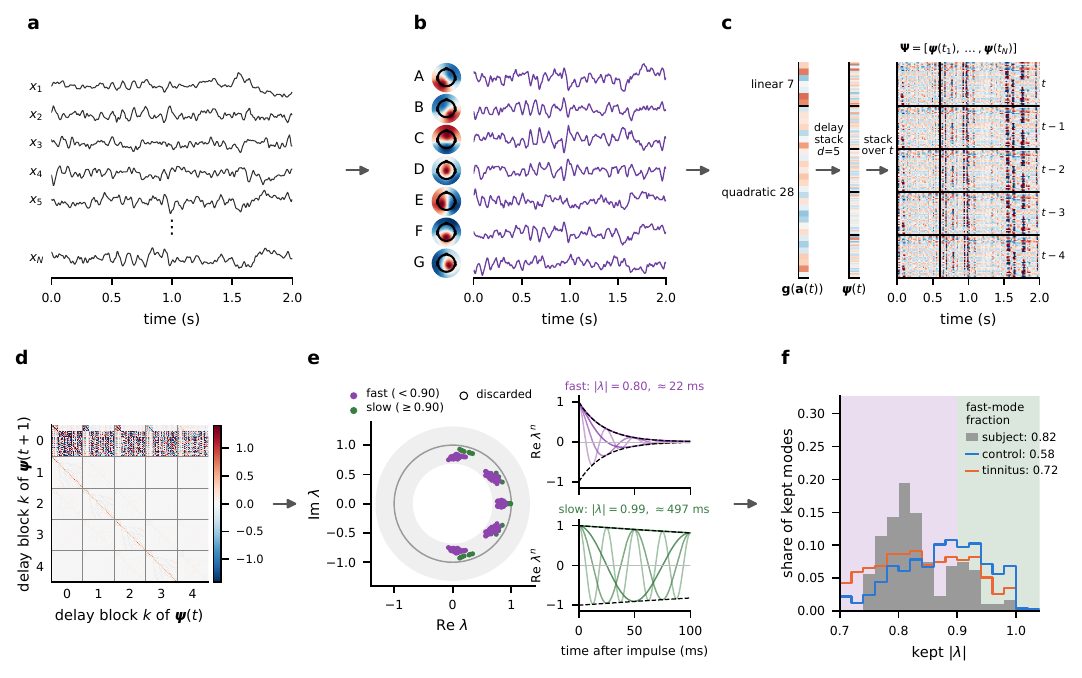}
\caption{\scriptsize Koopman pipeline, illustrated for one tinnitus subject (D2). (a) Resting EEG $\mathbf{x}(t)$ from $N$ channels (2~s shown). (b) Projection onto the seven template maps gives the microstate activations $\mathbf{a}(t)=\mathbf{M x}(t)$. (c) Lifting: the 7 activations and their 28 quadratic products form $\mathbf{g}(\mathbf{a}(t))$ (35 observables), delay-stacked with $d=5$ Hankel lags into $\boldsymbol\psi(t)$ (175 observables) and stacked over time into $\boldsymbol\Psi$ (row blocks: delays $t,\dots,t-4$; the time point shown is outlined). (d) The least-squares Koopman operator of this subject, in the same block order: shift blocks below the diagonal copy each delay one step back, and the top block row holds the learned dynamics, the prediction of $\mathbf{g}(\mathbf{a}(t+1))$ from all delays. Shown for a rank-89 truncated fit (singular values of $\boldsymbol\Psi$ above $10^{-4}$ of the largest). (e) Eigenvalues of $\mathbf{K}$: fast ($|\lam|\in[0.70,0.90)$, purple), slow (retained, $|\lam|\geq0.90$, green) and discarded ($\big||\lam|-1\big|>0.3$, open circles); the gray region is the retained range. Right: impulse responses $\mathrm{Re}\,\lam^n$ of illustrative fast and slow modes at matched frequencies (0, 10, 20 and 40~Hz; lighter: higher). (f) The fast-mode fraction: the subject's distribution of retained $|\lam|$ (grey bars) against the dataset's other control and tinnitus subjects, pooled by group.}
\label{fig:pipeline}
\end{figure*}

\subsection*{Tinnitus and control groups largely share microstate topographies}
\label{sec:res-maps}
Group differences in microstate statistics are only interpretable if tinnitus and control subjects express the same maps, so we first compared maps fit to each dataset (D1-D3), and to tinnitus and control groups separately, with the template \cite{custo_electroencephalographic_2017}. Maps fit to each whole dataset matched the template closely in D1 and D2 (mean $|r|=0.96$ in both), but less well in D3 (0.61; Fig.~\ref{fig:S-selffit}), consistent with that dataset's lower template Global Explained Variance (GEV). Maps fit separately to tinnitus and control subjects were highly similar to each other in D1 and D2 (mean $|r|=0.92$ in both; Fig.~\ref{fig:S-groupmaps}), and matching tinnitus to control maps directly reproduced the template labeling for 7/7 states. In D3, maps E and G differed between groups ($|r|=0.09$ and 0.32; direct pairing agreed for 5/7 states). Maps fit on one group explained nearly as much variance in the other group as in their own (between group GEV differences: 0.002--0.003 D1, 0.012--0.016 D2, 0.05--0.06 D3).

\subsection*{Spectral and microstate differences do not survive hearing adjustment, while the fast mode fraction does}
\label{sec:res-rep}
\begin{figure*}[!htbp]
\centering
% \figfile{fig9_forest_adjustment}{0.86\textwidth}
\includegraphics[width=0.86\textwidth]{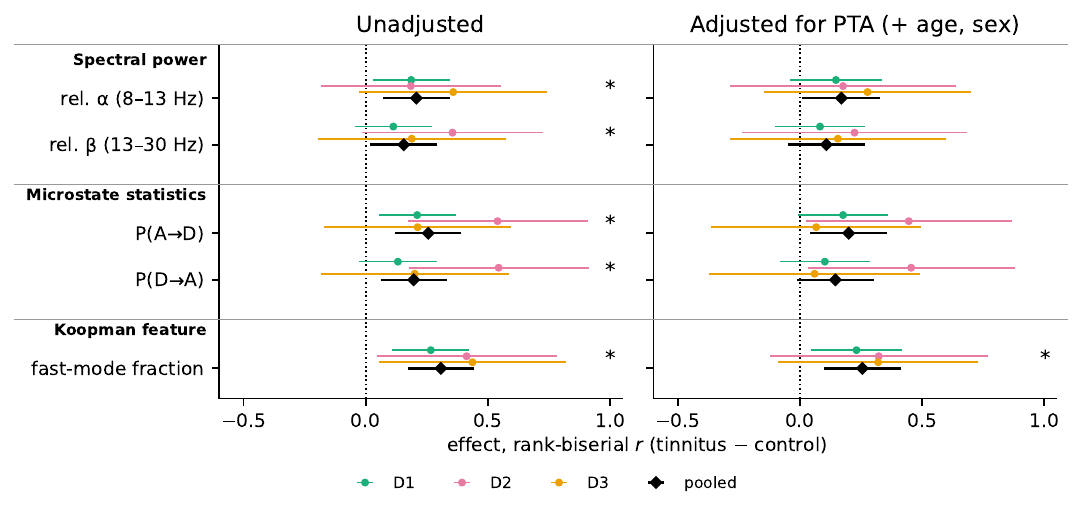}
\caption{\scriptsize Key features before (left) and after (right) adjustment for PTA (+ age, sex). Rows: spectral, microstate and Koopman statistics that replicate without adjustment. Colored points: per-dataset effect (rank-biserial $r$, positive = higher in tinnitus) with $\pm1.96$~SE; black diamonds: inverse-variance pooled effect; *: replicates (same sign in all datasets, pooled $q<0.05$).}
\label{fig:forest}
\end{figure*}

\begin{figure*}[!htbp]
\centering
% \figfile{fig10_feature_vs_pta_binned}{0.9\textwidth}
\includegraphics[width=0.9\textwidth]{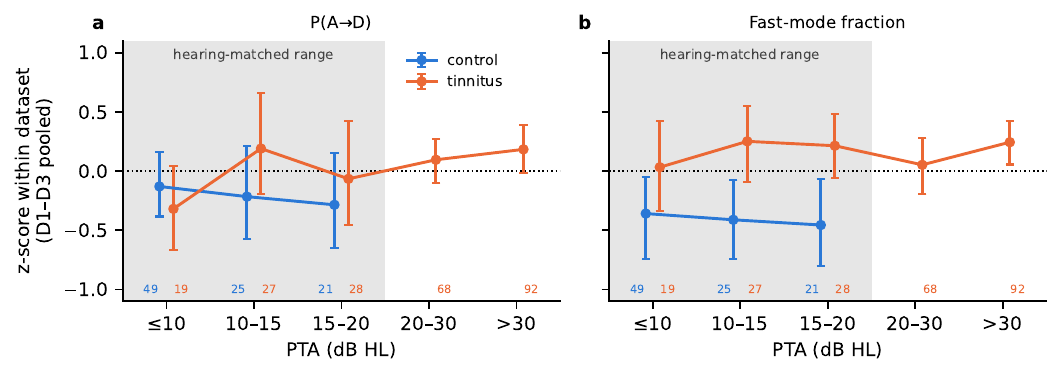}
\caption{\scriptsize Tinnitus and control groups compared at the same hearing level. (a) The transition probability $P(\mathrm{A}\to\mathrm{D})$, the strongest replicating microstate statistic; (b) the fast-mode fraction. Each feature is z-scored within its dataset and D1--D3 are pooled; points are group means per PTA bin with 95\% bootstrap CIs, shown for bins with at least 5 subjects of that group (numbers: subjects per bin). Shaded: bins with at least 5 subjects of both groups. Within the matched range $P(\mathrm{A}\to\mathrm{D})$ shows no consistent group difference, whereas the fast-mode fraction is higher in tinnitus at every matched hearing level.}
\label{fig:pta}
\end{figure*}

We next asked which features differ between tinnitus and control subjects consistently across the three datasets, and whether these differences remain once hearing level is accounted for. A feature was said to replicate if its effect had the same sign in all three datasets and was significant when pooled, after FDR correction (Methods). Table~\ref{tab:rep} summarizes the same replication test applied to the spectral baseline, microstate statistics, and Koopman fast-mode fraction (feature sets: Table \ref{tab:features}, Fig.~\ref{fig:S-rep}); Fig.~\ref{fig:forest} shows the key features before and after adjustment for hearing (plus age and sex where available), and Fig.~\ref{fig:pta} compares the groups at the same hearing level. 

\emph{Spectral power.} For comparison, relative $\alpha$ and $\beta$ power of the EEG (both, when averaged over the 15 common channels and on a channel-by-channel basis) replicated without adjustment but failed to replicate after PTA adjustment (Table~\ref{tab:rep}, Fig \ref{fig:forest}). A cluster based permutation test across neighboring channels (Methods) showed that relative $\alpha$ was higher in tinnitus at every channel, but not significantly so after FDR correction across the six spectral features ($q=0.14$; Figs.~\ref{fig:S-topo} and~\ref{fig:S-alpha}).

\emph{Microstate statistics.} No statistic differed between groups within any single dataset after FDR correction (Figs.~\ref{fig:S-box} and~\ref{fig:S-msdyn}), and per-feature effect sizes agreed only weakly between datasets (Spearman $\rho=0.07$--0.26). Pooled across datasets, \prov{2 of 56} statistics replicated (Table~\ref{tab:S-ms}), both transitions between maps A and D, more frequent in tinnitus ($P(\mathrm{A}\to\mathrm{D})$: $\rrb=$ \prov{0.21 / 0.54 / 0.21} in D1 / D2 / D3). After adjustment for PTA (and age, sex), the two pooled effects shrank by \prov{22\%} and \prov{26\%} and \prov{none} of the 56 statistics replicated (all $q\geq$ \prov{0.88}), although the pooled interval of $P(\mathrm{A}\to\mathrm{D})$ still excluded zero before correction (Fig.~\ref{fig:forest}). The between-dataset agreement of effect sizes fell to \prov{$\rho=-0.11$--0.05}. Consistent with a hearing signature, the microstate statistics predicted PTA within the largest dataset D1's tinnitus group (cross-validated $\rho=0.19$, $p=0.005$; Table~\ref{tab:S-pta}). 

\begin{figure*}[!htbp]
\centering
% \figfile{fig11_fast_mode_composition}{\textwidth}
\includegraphics[width=\textwidth]{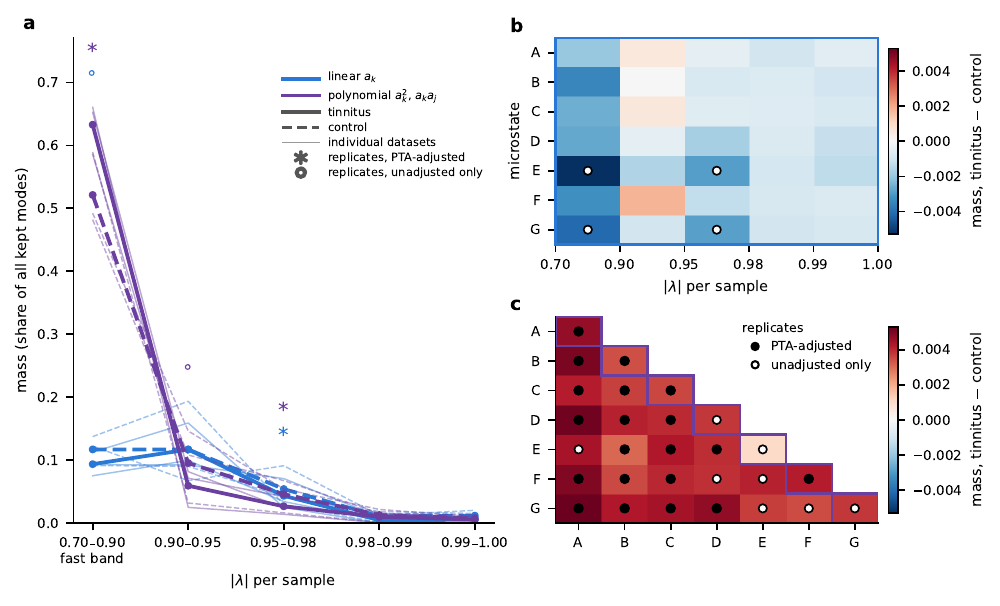}
\caption{\scriptsize Where the tinnitus difference in the Koopman spectrum lies. Mass: eigenvector energy share summed over a $|\lam|$ band's modes, divided by all retained modes (so all masses sum to 1); eigenvectors of the operator on unit-variance observables, so shares do not depend on the observables' units; datasets weighted equally. (a) Mass of linear terms and of polynomial terms (squares and cross terms) by $|\lam|$ *: the class difference replicates after PTA adjustment, $^\circ$: without adjustment only. (b) Tinnitus minus control linear mass per microstate and $|\lam|$ band. (c) Tinnitus minus control polynomial mass in the fast band: squares $a_k^2$ on the outlined diagonal, cross terms $a_ka_j$ below. Filled dots: replicate after PTA adjustment; open dots: without adjustment only. Replicates: same sign in all three datasets and pooled BH $q<0.05$.}
\label{fig:composition}
\end{figure*}

\emph{Koopman fast-mode fraction.} The fraction of fast-decaying modes was higher in tinnitus in every dataset ($\rrb=$ \prov{0.27 / 0.41 / 0.44}; pooled $r=$ \prov{0.31}, $p=$ \prov{$9\times10^{-6}$}) and remained so after adjustment for PTA (and age, sex): \prov{0.23 / 0.32 / 0.32}, pooled $r=$ \prov{0.26}, $p=$ \prov{0.002}, a shrinkage of \prov{17\%} (Fig.~\ref{fig:forest}). The adjusted effect was individually significant only in D1 ($p=$ \prov{0.015}); D2 and D3 contribute through pooling. The excess is mirrored by a deficit of modes with $|\lam|\in[0.95,0.98)$. Both are one shift from moderate to fast decay, carried by the polynomial terms (Fig.~\ref{fig:composition}a), and conditional on the fast-mode fraction the moderate-decay bands carry no further group information (Table~\ref{tab:S-beyond}). Unlike the microstate statistics, the Koopman $|\lam|$ quantiles did not predict PTA within D1's tinnitus group ($p=0.13$; Table~\ref{tab:S-pta}). No other Koopman feature carried group information beyond the fast-mode fraction (Table~\ref{tab:S-perfeat-koop}): of \prov{60} testable candidates (the $|\lam|$ quantiles, other $|\lam|$ bands, the mode-frequency distribution, $|\lam|\times$mode-frequency cells, the composition of fast modes and amplitude-weighted measures), \prov{24} replicated after PTA adjustment but \prov{none} after additionally adjusting for the fast-mode fraction (Table~\ref{tab:S-beyond}).

\emph{Sensitivity to the pooling model.} Pooling with a random-effects model instead of Stouffer's method (Methods) left the main results unchanged. The fast-mode fraction replicated before and after PTA adjustment, and no microstate statistic or spectral feature did after adjustment (Table~\ref{tab:S-re}). Only borderline unadjusted results changed, such as which microstate transitions replicated.

\begin{table}[!htbp]
\caption{Feature-level replication across D1, D2 and D3.}
\label{tab:rep}
\centering
\footnotesize
\setlength{\tabcolsep}{3pt}
\begin{tabular}{lcc}
\toprule
Feature set & Unadjusted & PTA-adjusted \\
\midrule
Spectral power, 1/$f$ slope (6; channel-wise)  & \prov{2 / 6}   & \prov{0 / 6} \\
Microstate statistics (56)       & \prov{2 / 56}  & \prov{0 / 56} \\
Koopman fast-mode fraction (1)   & \prov{yes} ($r=$ \prov{0.31}) & \prov{yes} ($r=$ \prov{0.26}) \\
\bottomrule
\end{tabular}
\\[2pt]\raggedright\footnotesize Number of features that replicate: same sign in all three datasets and
pooled Stouffer BH $q<0.05$ within the set (same sign by chance: 25\%). $r$: inverse-variance pooled
rank-biserial effect. Spectral replicating features (unadjusted): relative $\alpha$ and $\beta$ power.
\end{table}

\begin{figure*}[!htbp]
\centering
% \figfile{fig7a_lambda_quantile_differences}{0.85\textwidth}
\includegraphics[width=0.85\textwidth]{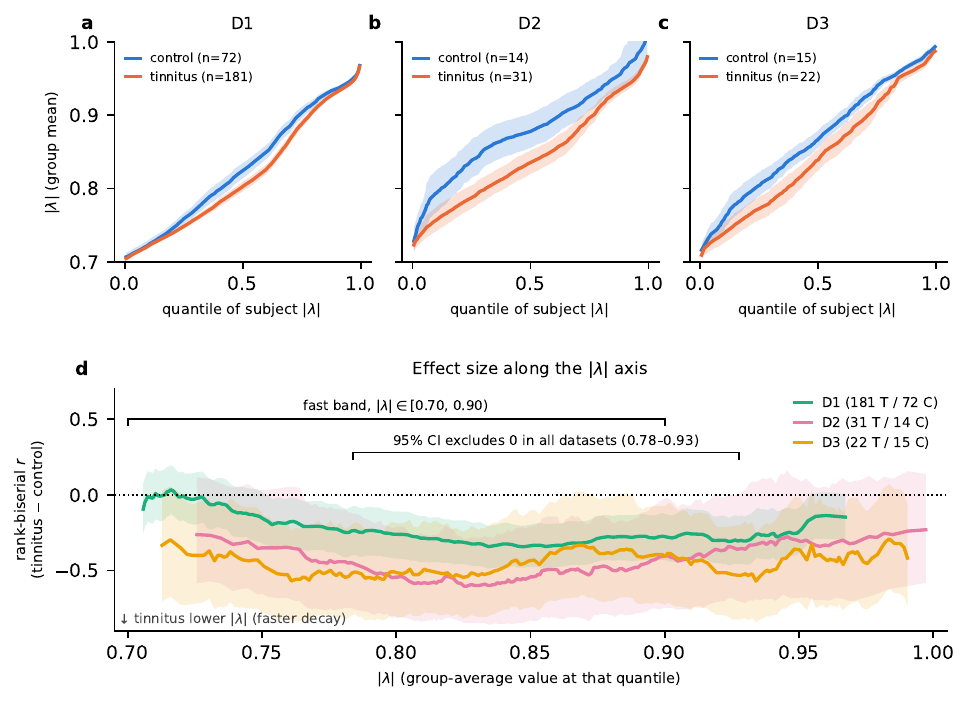}
\caption{\scriptsize Retained $|\lam|$ distributions. (a)--(c) Mean quantile function of each subject's retained $|\lam|$, per group and dataset. (d) Group effect ($\rrb$, tinnitus minus control) at each quantile, plotted against the group-average $|\lam|$ at that quantile; negative values: lower $|\lam|$ (faster decay) in tinnitus. Brackets: the fast band, and the $|\lam|$ range (quantiles 0.24--0.80, mapped to $|\lam|$ by averaging over datasets) where the 95\% CI excludes zero in all three datasets. Shaded: 95\% bootstrap intervals (1000 resamples of subjects within group).}
\label{fig:lambda}
\end{figure*}

\subsection*{The replicating Koopman effect: faster-decaying modes}
\label{sec:res-koop}
Each subject's Koopman operator yields a set of eigenvalues whose magnitude $|\lam|$ sets how quickly a mode decays (Fig.~\ref{fig:pipeline}e). To see where in this spectrum the groups differ, without committing to a band in advance, we compared the full distribution of retained $|\lam|$. Tinnitus subjects had lower $|\lam|$ across most of the retained spectrum in all three datasets (Fig.~\ref{fig:lambda}): the group difference had the same sign at 97\% of quantiles, and its 95\% CI excluded zero in every dataset for quantiles 0.24--0.80 ($|\lam|\approx0.78$--0.93). Equivalently, the fraction of eigenvalues with $|\lam|\in[0.70,0.90)$, modes with time constants of roughly 14--47~ms, was higher in tinnitus ($\rrb=$ \prov{0.27 / 0.41 / 0.44}; Stouffer \prov{$z=4.4$}); this band defines the fast-mode fraction. Chosen without the dataset it was tested on, the band was similar ($[0.70,0.81)$ to $[0.70,0.84)$) and the held-out effect was positive in every dataset after PTA adjustment ($r=$ 0.27 / 0.28 / 0.36 in D1 / D2 / D3); combined across the three held-out tests, $z=3.40$, $p=0.0003$ (Table~\ref{tab:S-lodo}). The effect spanned all mode frequencies (Fig.~\ref{fig:S-lamfreq}), which relate to the embedded dimensions through their respective eigenvectors.

\subsection*{What the fast-decaying modes are}
\label{sec:res-composition}
Because each mode combines the 35 observables and their delays, its eigenvector shows which functions of the microstate activations it describes. These can be the activations themselves (linear terms), map strength (squares) or co-activation (cross terms). We used this to ask what the fast-decaying modes represent, and whether they re-express microstate statistics or the power spectrum.

\emph{They are mainly dynamics of map strength and co-activation.} Each mode's eigenvector was computed for the operator on unit-variance observables ($\mathbf{D}\mathbf{K}\mathbf{D}^{-1}$, $\mathbf{D}$ the inverse standard deviations; same eigenvalues), so that its energy shares do not depend on the observables' units. Fast modes carried \prov{13\%} of their eigenvector energy on the linear activations, \prov{21\%} on squared and \prov{66\%} on cross terms (if all 35 observables contributed equally: 20\%, 20\% and 60\%), whereas slow modes ($|\lam|\geq0.90$) carried \prov{68\%} on the linear activations (Table~\ref{tab:S-decomp}). This composition is expected algebraically. A nonlinear process described by the co-activation of two states, embedded as $a_ka_l$, is displayed by an eigenvector with a weighting on that co-activation and associated eigenvalue $\lambda_{ij}$. The eigenvalues on the products $a_ka_l$ may represent the pairwise products $\lam_i\lam_j$ of the linear eigenvalues $a_k$ and $a_l$ as well. Products of slow linear modes ($|\lam|\in[0.90,0.98)$) fall at $|\lam|\in[0.81,0.96)$, largely inside the fast band. The linear share of fast modes stayed below \prov{0.36} at every mode frequency and was near zero above 45~Hz (Fig.~\ref{fig:S-linfreq}), but the large share of polynomial share of fast modes might be attributed to linear dynamics through this association. To identify whether these differences are from products of slow linear modes or from modes of higher order polynomial embeddings, we compared the fast mode fraction with phase randomized surrogates of the same activation signal. These surrogates preserved the first order auto and cross spectra but did not contain any of the higher order temporal structure. Surrogates reproduced the fast mode fraction results with a similar PTA-adjusted group effect (pooled $r=$ 0.23, $p=$ 0.007, vs.\ 0.26 on the original data, Table~\ref{tab:S-surr}), showing that the fast-mode fraction is therefore a function of the second-order (cross-spectral) structure of the activations.

\emph{Where the class difference lies.} Tinnitus subjects had more modes with $|\lam|\in[0.70,0.90)$ (adjusted $r=$ \prov{0.26}) and fewer with $|\lam|\in[0.90,0.98)$ ($r=$ \prov{$-0.20$} and \prov{$-0.24$}), with no difference near the unit circle. Modes that decay moderately in controls (time constants of about 50--250~ms) decay fast in tinnitus. This shift was carried mainly by the polynomial terms (squares and cross terms together), as expected when the slow linear modes themselves shift towards faster decay since their products land in the fast band. Their mass was higher in tinnitus in the fast band (adjusted $r=$ \prov{0.27}; tinnitus \prov{0.63} vs.\ control \prov{0.52}, higher in every dataset) and lower in the two bands below it ($r=$ \prov{$-0.15$} and \prov{$-0.19$}; the latter replicating after PTA adjustment, the former without adjustment only). Linear mass was, if anything, lower in tinnitus (fast band: $r=$ \prov{$-0.21$}, replicating without adjustment only; $|\lam|\in[0.95,0.98)$: \prov{$-0.19$}, replicating after adjustment Fig.~\ref{fig:composition}a). Again, with the shifting of these linear term products, these contribute to the fast band through the polynomial terms, even though no single microstate's linear terms replicated after adjustment (\prov{4/35} without, maps E and G; Fig.~\ref{fig:composition}b). Within the fast band the polynomial excess was broad. All \prov{28} squares and pairs replicated without adjustment and \prov{20} after PTA adjustment; those that did not mostly involve maps E, F and G (Fig.~\ref{fig:composition}c). The \emph{composition} of fast modes also shifted slightly toward cross terms in tinnitus (mean cross-term share, adjusted $r=$ \prov{0.28}, replicating): tinnitus subjects have more fast map-strength and co-activation modes, and their fast modes lean slightly more on co-activation.

\emph{Relation to microstate statistics and to the power spectrum.} The fast-mode fraction correlated with microstate summaries inconsistently across datasets (mean segment duration: $\rho=$ \prov{$+0.66$ / $-0.37$ / $-0.03$}; Table~\ref{tab:S-corr}), and adjusting for them did not change the group effect (Table~\ref{tab:explain}). It correlated consistently with the relative $\alpha$ power of the scalp averaged EEG ($\rho=$ 0.70 / 0.77 / 0.38 in D1 / D2 / D3, channel mean over the same 60~s window) and negatively with relative $\delta$ power. Relative $\alpha$ and $\beta$ power themselves differed between groups, but, like the microstate statistics, none replicated after PTA adjustment (all $q\geq$ \prov{0.25}). Adjusting the fast-mode fraction for relative $\alpha$ power reduced the pooled effect from \prov{0.26} to \prov{0.17} ($p=$ \prov{0.009}), and for all five bands to \prov{0.11} ($p=$ 0.036): much of the hearing-adjusted Koopman effect is shared with the spectral $\alpha$/$\delta$ balance, which ties the fast modes to a familiar EEG measure, although the spectral features themselves did not differ between groups after hearing adjustment.

\begin{table}[!htbp]
\caption{Group effect on the fast-mode fraction after adjusting for candidate explanations.}
\label{tab:explain}
\centering
\footnotesize
\setlength{\tabcolsep}{3pt}
\begin{tabular}{lcc}
\toprule
Model (all include PTA + age, sex) & Pooled $r$ & $p$ \\
\midrule
PTA only                          & \prov{0.26} & \prov{0.002} \\
+ mean segment duration           & \prov{0.25} & \prov{0.001} \\
+ segment rate                    & \prov{0.26} & \prov{0.001} \\
+ relative $\delta$ power         & \prov{0.18} & \prov{0.014} \\
+ relative $\alpha$ power         & \prov{0.17} & \prov{0.009} \\
+ relative $\beta$ power          & \prov{0.27} & \prov{0.002} \\
+ relative power, 5 bands         & \prov{0.11} & \prov{0.036} \\
+ 1/$f$ slope                     & \prov{0.24} & \prov{0.002} \\
\bottomrule
\end{tabular}
\\[2pt]\raggedright\footnotesize Pooled $r$: inverse-variance mean of the per-dataset adjusted effects;
$p$: Stouffer.
\end{table}

\subsection*{Robustness to embedding and fitting choices}
\label{sec:res-robust}
The fast-mode fraction depends on choices made when fitting the operator. The Hankel embedding dimension $d$ and delay $\tau$, and whether the least-squares fit is regularized. We therefore tested whether the effect depends on these choices. With consecutive-sample embedding ($\tau=1$), the fast-mode fraction replicated at every embedding dimension (Stouffer \prov{$z=4.4$, 2.9, 3.0} for $d=5$, 10, 15; Fig.~\ref{fig:S-robust}). Across all 12 embeddings it replicated in \prov{6}; D2 showed the effect in all 12 and D3 in 11, whereas in D1 it reversed at $\tau=2$ and 4 for $d\ge10$ and at $\tau=4$ for $d=5$. Regularizing the fit with a truncated SVD left the
hearing-adjusted effect intact ($r=$ \prov{0.25--0.29}, $p\leq$ \prov{0.005} at every threshold) and removed the per-subject instability of the untruncated fit (Table~\ref{tab:S-trunc}). When $d$ and $\tau$ were instead selected without the held-out dataset (by mean cross-dataset AUROC on the other two), $d=5$, $\tau=1$ was chosen in all three folds. The group difference lay in the eigenvalue magnitudes rather than their phases. The mode frequencies at which tinnitus subjects had excess fast-decaying modes differed between datasets (Fig.~\ref{fig:S-cloud}), and the distance between datasets relative to that between groups was \prov{2.3} for eigenvalue phase against \prov{1.4} for magnitude (Table~\ref{tab:S-cloud}).

\subsection*{Classification and hearing level}
\label{sec:res-clf}
Finally, we asked whether these features classify individual subjects when a classifier trained on one dataset is applied to another, and how this compares with hearing level alone. The Koopman classifier, which uses each subject's full distribution of retained $|\lambda|$ (Methods), also supported cross-dataset classification (Table~\ref{tab:clf}): mean AUROC 0.70 (range 0.60--0.80; above chance at $p<0.05$ in 6/6 directions) and mean MCC 0.21 (worst direction 0.09), against mean AUROC 0.46--0.63 for the four microstate classifiers (Table~\ref{tab:S-clf}). Koopman exceeded the linear-SVM baseline in all six directions on every metric (paired Wilcoxon $p=0.031$, the minimum attainable with six pairs) and the RBF SVM in 5/6 directions by AUROC (ROC curves: Fig.~\ref{fig:S-roc}). Across Hankel embeddings, the size of the fast-mode fraction effect correlated with cross-dataset AUROC (\prov{$\rho=0.64$, $p=0.03$}).

Hearing level alone separated the groups (AUROC 0.78--0.90; Table~\ref{tab:hearing}). This reflects how the cohorts were recruited; tinnitus subjects with hearing loss due to the high co-occurrence and controls with near-normal hearing. PTA measures hearing, and cannot identify tinnitus in subjects with normal hearing and would not separate the groups if hearing were matched. It therefore quantifies the confound that an EEG marker of tinnitus must overcome, not a competing feature. In PTA-matched subsets, cross-dataset Koopman scores still separated the groups in D2 (7 vs.\ 7; AUROC 0.76) and D3 (10 vs.\ 10; 0.70) but only weakly in D1 (50 vs.\ 50; 0.56, $p\approx0.13$), stable across 200 random matchings (Fig.~\ref{fig:S-matched}); the pooled subject-weighted AUROC was 0.57 [0.47, 0.68] ($p=0.10$), whereas the microstate SVM fell to or below chance (0.45 / 0.13 / 0.41; pooled 0.45). Neither feature set's scores correlated with tinnitus severity after FDR correction (all $q\geq0.46$; Table~\ref{tab:S-sev}; nominally only the within-dataset microstate scores in D3, $\rho=0.48$, $p=0.025$).

\begin{table}[!htbp]
\caption{Cross-dataset classification (train $\to$ test).}
\label{tab:clf}
\centering
\footnotesize
\setlength{\tabcolsep}{2.5pt}
\begin{tabular}{lcccccc}
\toprule
 & \multicolumn{3}{c}{AUROC} & \multicolumn{3}{c}{MCC} \\
\cmidrule(lr){2-4}\cmidrule(lr){5-7}
Direction & Lin.\ SVM & RBF SVM & Koopman & Lin.\ SVM & RBF SVM & Koopman \\
\midrule
D1 $\to$ D2 & 0.28 & 0.62 & 0.70 [0.53, 0.85] & $-0.41$ & 0.21    & 0.22 \\
D1 $\to$ D3 & 0.53 & 0.47 & 0.78 [0.61, 0.92] & $-0.12$ & 0.00    & 0.29 \\
D2 $\to$ D1 & 0.47 & 0.60 & 0.63 [0.54, 0.70] & 0.01    & 0.10    & 0.19 \\
D2 $\to$ D3 & 0.48 & 0.74 & 0.69 [0.51, 0.85] & 0.05    & 0.25    & 0.10 \\
D3 $\to$ D1 & 0.45 & 0.57 & 0.60 [0.52, 0.67] & $-0.06$ & $-0.04$ & 0.09 \\
D3 $\to$ D2 & 0.57 & 0.78 & 0.80 [0.65, 0.94] & 0.01    & 0.32    & 0.36 \\
\midrule
Mean  & 0.46 & 0.63 & 0.70 & $-0.08$ & 0.14    & 0.21 \\
Worst & 0.28 & 0.47 & 0.60 & $-0.41$ & $-0.04$ & 0.09 \\
\bottomrule
\end{tabular}
\\[2pt]\raggedright\footnotesize Microstate baselines use the 56 statistics; all four microstate classifiers: Table~\ref{tab:S-clf}. Koopman AUROC: 95\% bootstrap CI. MCC at the default threshold. Within-dataset 5-fold CV AUROC (D1 / D2 / D3): Koopman 0.61 / 0.76 / 0.73, linear SVM 0.57 / 0.64 / 0.54.
\end{table}

\begin{table}[!htbp]
\caption{Hearing level and EEG scores.}
\label{tab:hearing}
\centering
\footnotesize
\setlength{\tabcolsep}{3pt}
\begin{tabular}{lccc}
\toprule
 & D1 & D2 & D3 \\
\midrule
PTA AUROC (group)                       & 0.90 & 0.78 & 0.83 \\
Koopman AUROC, within-CV                & 0.61 & 0.76 & 0.73 \\
PTA-matched pairs                       & 50   & 7    & 10   \\
Matched AUROC, Koopman (cross-dataset)  & 0.56 & 0.76 & 0.70 \\
Matched AUROC, microstate SVM           & 0.45 & 0.13 & 0.41 \\
\bottomrule
\end{tabular}
\\[2pt]\raggedright\footnotesize Matched AUROC: scores from classifiers trained on the other datasets, combined by within-dataset rank averaging. Pooled AUROC by stratified Mann–Whitney (U summed over datasets, divided by the total number of within-dataset tinnitus–control pairs); p by permutation within dataset: Koopman 0.57 ($p=0.10$), microstate 0.45. PTA AUROC reflects cohort composition (tinnitus subjects with hearing loss, controls with near-normal hearing) and is shown as the size of the hearing confound, not as a tinnitus marker.
\end{table}

% =====================================================================

\section*{Discussion}
Across three independent resting-state EEG datasets, tinnitus and control subjects largely shared microstate topographies, and the microstate statistics that differed between them were partly explained by hearing level. A Koopman-operator description of the same microstate signal revealed a shift toward faster-decaying dynamics in tinnitus that was consistent across datasets and persisted after adjustment for hearing level; it also supported cross-dataset classification, with a mean AUROC of 0.7 across all train-test directions. Maps fit separately to each group were nearly identical in datasets D1 and D2, and matched for five out of the seven states in D3. Clustered microstates showed differences from the template and between groups for D3. This was partly due to a few channels, T7, T8 and O2, whose signal the template does not describe (Fig.~\ref{fig:S-noisychannels}).

While the clustered microstates were comparable between groups, they showed temporal differences in the presentation of these states. This difference was captured in the temporal dynamics in A an D transitions when pooled across datasets. This specific microstate difference has been previously reported in tinnitus and hearing loss \cite{chen_resting-state_2026, raeisi_enhanced_2025}. The Koopman linear terms showed the same hearing dependence (Fig.~\ref{fig:composition}: the linear mass of maps E and G differed before, but not after, PTA adjustment). Significant differences have been shown with reference to the time coverage of microstate G in hearing loss alone \cite{cai_altered_2019}. Because tinnitus and hearing loss are strongly associated, both clinically and in these datasets (PTA alone separated the groups with AUROC 0.78--0.90), group differences in microstate statistics cannot be attributed to tinnitus without accounting for hearing. Here they appear to be partly confounded by hearing level. After adjustment for PTA the pooled effects shrank by about a quarter and no statistic replicated, the replicating transition $P(\mathrm{A}\to\mathrm{D})$ increased with PTA within the tinnitus group (pooled Spearman $z=1.99$, $p=0.047$; Fig.~\ref{fig:S-ptascatter}), and microstate statistics predicted PTA within the tinnitus group of D1, the largest dataset ($\rho=0.19$, $p=0.005$), whereas neither the fast-mode fraction ($p=0.94$) nor the Koopman $|\lambda|$ quantiles ($p=0.13$) did; D2 and D3 were too small for this cross-validated prediction. This is consistent with the central gain model of tinnitus \cite{norena_integrative_2011}, in which central changes follow peripheral hearing loss.

Prior literature for EEG differences between tinnitus and control has used spectral comparisons to show changes in alpha and delta power in tinnitus vs control \cite{weisz_tinnitus_2005, vanneste_neural_2010}. In our study, relative band power of the scalp averaged EEG showed that tinnitus subjects had higher relative $\alpha$ and $\beta$ power, and $\gamma$ power did not differ consistently across datasets. The same pattern held at temporal channels T7/T8 and channel by channel, but none of the spectral differences survived adjustment for hearing loss. Like the microstate statistics, these differences appear to track hearing level (as shown in previous studies \cite{pierzycki_whole_2016}). The Koopman operator results, however, are not independent of these spectral differences. As shown in Table \ref{tab:explain}, adjusting for $\delta$ and $\alpha$ power reduced the hearing-adjusted group effect on the fast-mode fraction (pooled $r$ from \prov{0.26} to 0.18 and 0.17), so much of the effect is shared with the $\alpha$/$\delta$ balance. Unlike relative band power, the fast-mode fraction retained a group difference after hearing adjustment.

Microstate statistics come from a Markov chain on the winning map, whose transition matrix is itself a transfer operator restricted to $k=7$ indicator functions, with mean duration arising from its diagonal; yet microstate sequences are not Markovian \cite{von_wegner_information-theoretical_2018}. The Koopman operator used here acts on a richer set of functions of the same microstate activations: (i) the continuous activation of every map at every time point instead of a winner-take-all label, since transitions are gradual rather than abrupt and, at low global field power, the field is shared across several templates \cite{mishra_eeg_2020,dinov_modeling_2017}; and (ii) their delays and products, whose dynamics are the activations' second moments, the correlation function of a product $a_ka_l$. Here, the tinnitus difference lies mainly in the quadratic part. Tinnitus subjects have more modes in which map strengths and co-activations decay within roughly 14--47~ms, spread across maps rather than tied to one microstate. The winner-take-all sequence discards the continuous activations and their timing, and accordingly microstate features do not survive PTA adjustment in their effect. The five band power features, which summarize the same spectra more coarsely accounted for part of it (Table~\ref{tab:explain}). The group difference in eigenvalues lies primarily in the distribution of their magnitude, with tinnitus subjects exhibiting more fast-decaying modes, while the distribution of their phase (frequency) shows less consistent group-level differences. Of the eigenvalue phase features, only the fractions of modes at 13--30~Hz and 30--50~Hz replicated after PTA adjustment (Stouffer $z=$ \prov{3.01} and \prov{$-2.52$}), and both were strongly correlated with the fast-mode fraction. These features could be interpreted as a by-product of the shift in decay dynamics rather than independent group-level differences.

While the microstates themselves represent discrete configurations that are relatively stable for 50 to 120 milliseconds \cite{kleinert_reliability_2024}, durations of specific states have shown interesting group differences for various pathologies. As the fast modes decay within \prov{14--47}~ms, they are faster than the microstate segments themselves. These shorter fluctuations are distinct from previous microstate studies in tinnitus showing significantly higher microstate occurrence rates \cite{raeisi_enhanced_2025} and longer duration of microstates \cite{cao_microstate_2020} showing that the tinnitus group needed more time to maintain steady states for mental processes \cite{rieger_15_2016,lehmann_eeg_2005}. In our data, adjusting for mean segment duration and segment rate did not change the group effect on the fast-mode fraction (Table~\ref{tab:explain}). Under the central gain model, tinnitus emerges when the increase in central gain that follows peripheral deafferentation is not balanced, increasing spontaneous activity and excitability \cite{norena_integrative_2011}. A more excitable network could also shorten the correlation time of map strength and co-activation. Because PTA does not capture all deafferentation, such as cochlear synaptopathy in listeners with normal audiograms \cite{schaette_tinnitus_2011}, the effect that remains after hearing adjustment could also partly reflect hearing damage not measured by PTA.

Fast-decaying modes may also reflect irregular, unstructured variance in map strength and co-expression rather than a single dynamical process, fluctuations that are not predictable from the preceding \prov{$\sim$20}~ms of the embedded activations. Measurement noise is unlikely to explain the group difference; the effect had the same sign in three datasets recorded with different systems, was unchanged after adjusting for the 1/$f$ slope of the spectral power, which flattens with broadband noise and muscle activity, while $\gamma$ power, the band most affected by muscle artifact, did not differ between groups (Table~\ref{tab:explain}), and held when a truncated SVD was applied to the operator, thresholding the noise (Table~\ref{tab:S-trunc}). The fast-mode fraction was also higher in subjects with more relative $\alpha$ power, a structured rhythm, rather than in those with flatter, noisier spectra. The fast modes could therefore point to neural fluctuations in temporal dynamics that are less predictable in tinnitus, which correlates with previous studies showing high classification accuracy between tinnitus and control using sample entropy \cite{jianbiao_eeg_2023, raeisi_enhanced_2025,sadeghijam_shannon_2021} and an increased chaotic nature.

The fast-mode fraction held when its embedding and band were chosen without the test dataset and at every embedding dimension with $\tau=1$ (Results). Because the fast-mode fraction is a one-dimensional summary of eigenvalue magnitudes, it needs no alignment of modes between subjects or datasets, and the eigenvalues are unchanged by rescaling the observables, so while dataset differences in channel count or signal amplitude may change the activation magnitude, the eigenvalue magnitudes remain stable. We also showed that the phase-based distribution varied more across datasets than between groups, showing that frequency-based features would be more vulnerable to a dataset shift. While tinnitus subjects show more fast-decaying modes than control subjects, the frequencies that display this shift vary between the datasets (Fig.~\ref{fig:S-cloud}).

The Koopman classifier performed better overall in cross-dataset classification than the microstate statistics (Table~\ref{tab:clf}), and the fast-mode fraction remained higher in tinnitus across the hearing-matched range (Fig.~\ref{fig:pta}). This classification performance, however, needs further validation before application as a diagnostic test for tinnitus. PTA alone separated the groups (Table~\ref{tab:hearing}), but as a measure of the hearing confound rather than of tinnitus. In PTA-matched pairs, where hearing cannot contribute, the Koopman scores remained above chance in all three datasets, whereas the microstate classifier fell to or below chance (0.45 / 0.13 / 0.41). Although the fast-mode fraction demonstrates a hearing-independent group difference that replicates across datasets, it is better suited as a group-level marker than as an individual classifier. This suggests that while the tinnitus percept is highly correlated with hearing loss, the fast-mode fraction provides a group difference beyond hearing level that, once related to tinnitus severity and treatment outcomes, could be used as a stratification measure for studies of the tinnitus percept.

These results further underline the necessity of testing EEG markers of tinnitus across independent sites and accounting for hearing loss before attributing them to the tinnitus percept. In our data, neither the microstate statistics nor the band power that differed between the groups survived adjustment for hearing level. Future EEG studies of tinnitus should recruit hearing-matched control groups or, at minimum, measure hearing and include it as a covariate, and validate candidate markers on independent datasets.

% =====================================================================
These findings have limitations: with no group with hearing loss but without tinnitus, hearing adjustment is statistical and, because controls had near-normal hearing, relies on extrapolation at hearing levels where few controls were recorded (Fig.~\ref{fig:pta}). Moreover, no association with tinnitus severity was found; severity scales differ across datasets, and the analysis is restricted to a binary group comparison that does not address subtypes. The physiological origin of the fast-decaying modes also remains open. Future work should test the fast-mode fraction, with its band and embedding fixed, on independent datasets that include hearing-loss subjects without tinnitus, and relate it to tinnitus severity and subtype.

\section*{Methods}
\label{sec:methods}
\subsection*{Datasets}
\label{sec:data}
Three publicly available resting-state EEG datasets were used (Table~\ref{tab:datasets}); we refer to
them as D1--D3 throughout.

\textbf{D1} \cite{wang_cross-subject_2023}: recordings from the Department of Otolaryngology, Sun Yat-sen Memorial Hospital, Sun Yat-sen University, acquired with a high-density 128-channel wet EEG system and a NetAmps 200 amplifier; 118 channels were retained (117 scalp electrodes and Cz) (9 electrodes removed by the original authors, and the E56 and E107 were removed after re-referencing). 14 subjects were excluded for having less than 60s of resting state data or mismatch in channel count. 

\textbf{D2} \cite{cuevas-romero_electroencephalography-based_2022}: a database of tinnitus subjects enrolled in an acoustic-therapy trial and healthy controls, recorded with a 16-channel headset and a g.USBamp amplifier; only the pre-therapy, eyes closed sessions were used. 1 subject was excluded for having less than 60s of resting state data.

\textbf{D3} \cite{torres-torres_characterization_2023}:  resting-state recordings of 22 tinnitus and 15 control subjects, acquired with a 27-channel system using standard 10--20 electrode names; electrode positions were not digitized, so the standard 10--20 montage was used. No subjects were excluded.

\begin{table}[!htbp]
\caption{Datasets after exclusions.}
\label{tab:datasets}
\centering
\footnotesize
\setlength{\tabcolsep}{2pt}
\begin{tabular}{lccccccc}
\toprule
 & $N$ & T / C & Ch. & Length (s) & PTA, T / C & Age, T / C & \% F, T / C \\
\midrule
D1 & 253 & 181 / 72 & 118 & $174 \pm 73$ & 25.5 / 10.5 & 45 / 40     & 59 / 54 \\
D2 & 45  & 31 / 14  & 16  & 181          & 24.4 / 8.4  & $^\dagger$  & 71 / 33$^\ddagger$ \\
D3 & 37  & 22 / 15  & 27  & 121          & 24.1 / 11.0 & n/a         & n/a \\
\bottomrule
\end{tabular}
\\[2pt]\raggedright\footnotesize T / C: tinnitus / control. PTA (dB HL) and age (years): group medians. \% F: percent female.
Severity: THI (D1), THI grade (D2), TFI (D3). All recordings resampled to 200~Hz.
$^\dagger$D2 records age only in three categories (18--29, 30--59, $\geq$60 years
\cite{cuevas-romero_electroencephalography-based_2022}): T 1 / 17 / 13, C 1 / 9 / 2 (2 controls missing).
$^\ddagger$Of 12 controls with recorded sex. n/a: not included in the D3 release.
\end{table}

\subsection*{Preprocessing}
\label{sec:prep}
All recordings were high-pass filtered at 0.1~Hz and cleaned with GEDAI \cite{ros_return_2025}, an artifact-removal method based on generalized eigendecomposition \cite{cohen_tutorial_2022} and designed for microstate analysis. The cleaned data were low-pass filtered at 50~Hz, resampled to 200~Hz, re-referenced to the common average and z-scored per channel. Electrode positions were set from standardized MNE montages \cite{gramfort_meg_2013}. The same cleaned data were used for both feature families. We used default GEDAI parameters, with an interpolated reference scheme and epoch size in cycles set to 12.

\subsection*{Microstate template and back-fitting}
\label{sec:template}
All analyses used the literature template of Custo et al. \cite{custo_electroencephalographic_2017} ($k=7$, maps A--G), spline-interpolated onto each dataset's montage. % (Fig.~\ref{fig:maps}). 
Because the maps are fixed without inference from the tinnitus datasets, within-dataset cross-validation involves no leakage between train and test subjects from map-fitting and map labels mean the same topography and polarity in every dataset. Interpolating each map back onto the template montage recovered it with $|r|\geq0.995$ in all datasets. The template explained 56\% (D1), 53\% (D2) and 41\% (D3) of the global variance (GEV; Table~\ref{tab:S-gev}). The lower GEV in D3 is partly due to specific channels that the template does not describe; the temporal channels T7 and T8 and O2 of D3 are essentially unexplained (Fig.~\ref{fig:S-noisychannels}). The use of a fixed template also allows for more comprehensive, overarching conclusions with reference to existing and emerging microstate literature \cite{koenig_eeg-meta-microstates_2024}.

To check that the template is representative, $k=7$ maps were also fit using the pycrostates package \cite{ferat_pycrostates_2022} (modified k-means on GFP peaks) to each dataset as a whole, as well as separately to its tinnitus and control subjects, and compared with the template by spatial correlation. These self-fit maps were used only for this comparison (Figs.~\ref{fig:S-selffit} and~\ref{fig:S-groupmaps}) to demonstrate that the template maps adequately capture the dominant topographies in each dataset.

\subsection*{Microstate statistics}
\label{sec:msfeat}
Microstate activations were calculated as the dot product of the EEG data with each template map, 
\begin{equation}
  a_k(t) = \mathbf{m}_k^\top \mathbf{x}(t), \quad k=1,\dots,7.
  \label{eq:activation}
\end{equation}
Each sample was assigned to the map with the largest absolute activation,
\begin{equation}
  s(t) = \arg\max_k |a_k(t)|, \quad t=1,\dots,T,
\end{equation}
giving the subject's microstate sequence. With $\delta_{i,j}$ the Kronecker delta ($1$ if $i=j$, $0$
otherwise), let $c(t)=1-\delta_{s(t),s(t-1)}$ mark the start of a new segment ($c(1)=1$). For each map $k$,
\begin{align}
  \text{Coverage}_k &= \frac{1}{T} \sum_{t=1}^{T} \delta_{s(t),k}, \label{eq:coverage}\\
  \text{Duration}_k &= \Delta t\,\frac{\sum_{t=1}^{T} \delta_{s(t),k}}{\sum_{t=1}^{T} \delta_{s(t),k}\, c(t)}, \label{eq:duration}
\end{align}
the fraction of samples assigned to map $k$ and the mean length of its segments ($\Delta t = 1/f_s$; in ms).
For each ordered pair of maps $k\neq l$,
\begin{equation}
  \text{Transition}_{k\to l} = \frac{\sum_{t=2}^{T} \delta_{s(t-1),k}\,\delta_{s(t),l}}{\sum_{t=2}^{T} c(t)}, \label{eq:transition}
\end{equation}
the fraction of all state changes that go from $k$ to $l$ (the 42 values sum to 1). With 7 coverages,
7 durations and 42 transitions, this gives 56 features. For Table~\ref{tab:explain} we also used two whole-sequence summaries: the mean segment duration across all maps, $T\Delta t / \sum_{t} c(t)$, and the segment rate, $\sum_{t} c(t) / (T\Delta t)$, the number of segments per second. Each of the features is independent of recording length, which differs across datasets. Microstate statistics were computed over each subject's full recording, whereas the Koopman and spectral features use a centered 60~s window.

\subsection*{Koopman operator features}
\label{sec:koopman}
With $\mathbf{a}(t)\in\mathbb{R}^7$ the microstate activations \eqref{eq:activation} of a centered 60~s window, the observables are the activations and all their quadratic monomials (7 linear terms and $7\cdot8/2=28$ products),
\begin{equation}
\begin{split}
  \mathbf{g}(\mathbf{a}) &= \big[\,a_1,\dots,a_7,\; a_1^2,\,a_1a_2,\dots,a_7^2\,\big]^\top\\
  &= \big[\,\{a_k\}_{k=1}^{7},\ \{a_k a_l\}_{1\le k\le l\le 7}\,\big]^\top \in \mathbb{R}^{35},
\end{split}
\label{eq:poly}
\end{equation}
lifted by a Hankel embedding \cite{takens_detecting_1981, arbabi_ergodic_2017} with $d$ lags of stride $\tau$,
\begin{equation}
\begin{split}
  \boldsymbol\psi(t) = \big[\,&\mathbf{g}(\mathbf{a}(t))^\top,\ \mathbf{g}(\mathbf{a}(t-\tau))^\top,\ \dots,\\
  &\mathbf{g}(\mathbf{a}(t-(d-1)\tau))^\top\,\big]^\top \in \mathbb{R}^{35d}.
\end{split}
\label{eq:hankel}
\end{equation}
A per-subject Koopman operator $\mathbf{K}$ was estimated by least squares on consecutive-sample snapshot pairs,
\begin{equation}
  \mathbf{K} = \boldsymbol\Psi_{+}\,\boldsymbol\Psi^{\dagger},
  \label{eq:koopman}
\end{equation}
where $\boldsymbol\Psi = [\boldsymbol\psi(t_1),\dots,\boldsymbol\psi(t_{N-1})]$ and
$\boldsymbol\Psi_{+} = [\boldsymbol\psi(t_2),\dots,\boldsymbol\psi(t_{N})]$, as in extended DMD \cite{williams_data_2015}. Each eigenvalue $\lam$ of $\mathbf{K}$ is a mode with per-sample persistence $|\lam|$ (decay time constant $-\Delta t/\ln|\lam|$) and frequency $f=|\arg\lam|\,f_s/2\pi$. Because the time step is one sample for every embedding, $|\lam|$ has the same meaning across embeddings. Eigenvalues with $\big||\lam|-1\big|\le0.3$ were retained. Quadratic observables produce sum frequencies, so retained modes can lie above the 50~Hz low-pass edge. Fig.~\ref{fig:pipeline} shows each step for one subject. For the PTA-prediction analysis, each subject's retained $|\lam|$ were also summarized by 20 quantiles and the fraction of eigenvalues retained ($|\lam|$ quantiles, 21 features).

We used the minimal Hankel embedding, $d=5$, $\tau=1$. It also had the highest mean AUROC when selecting parameters in a leave-one-dataset-out manner over a sweep of $d\in\{5,10,15\}$ and $\tau\in\{1,2,4,8\}$. The fast-mode effect itself replicated at every embedding dimension with $\tau=1$ (Fig.~\ref{fig:S-robust}), and the band was validated on held-out datasets (Table~\ref{tab:S-lodo}).

\subsection*{Spectral baseline}
\label{sec:spec}
As a conventional reference, relative band power and the aperiodic slope were computed from the cleaned EEG itself, as in other spectral studies of tinnitus. So that the same electrodes are compared across datasets, we used the 15 standard 10--20 electrodes recorded in all three (Fp1, Fp2, F7, F3, Fz, F4, F8, T7, C3, C4, T8, P7, P8, O1, O2); for D1's 128-channel net, the electrode nearest each 10--20 position (median 6$^\circ$, at most 12$^\circ$ away on the head) was used. For each channel, the power spectrum of the whole recording was estimated by Welch's method with 2~s segments; relative power was the share of 1--45~Hz power in the $\delta$ (1--4~Hz), $\theta$ (4--8~Hz), $\alpha$ (8--13~Hz), $\beta$ (13--30~Hz) and $\gamma$ (30--45~Hz) bands, and the 1/$f$ slope was the slope of a straight-line fit of log power against log frequency over 2--40~Hz. The six features were averaged over the 15 channels for the main spectral feature set, and were also tested at the temporal channels T7/T8 and at each channel separately (Benjamini--Hochberg correction over the 90 channel--feature pairs). As covariates for the fast-mode fraction (Table~\ref{tab:explain}), the same features were computed over the centered 60~s window used for the Koopman fit. Channel-level differences were also tested with a cluster-based permutation test \cite{maris_nonparametric_2007}. In each dataset, each channel's feature was compared between groups by the $t$-statistic of the group term of a linear model (with PTA, age and sex for the adjusted test), converted to a normal score, and the three datasets were combined by Stouffer's method. Neighboring channels (adjacency from the electrode positions) with $|Z|>1.96$ and the same sign formed clusters, scored by their summed $Z$. The largest cluster score was compared with its distribution over 1000 permutations of group labels within each dataset (for the adjusted test, permutation of the residuals of the covariate-only model, Freedman--Lane). The cluster $p$-values were corrected across the six spectral features with the Benjamini--Hochberg procedure.

\begin{table}[!htbp]
\caption{Feature sets and where they are used.}
\label{tab:features}
\centering
\footnotesize
\setlength{\tabcolsep}{3pt}
\begin{tabular}{p{0.24\linewidth}cp{0.38\linewidth}p{0.26\linewidth}}
\toprule
Feature set & $n$ & Definition & Used for \\
\midrule
Microstate statistics & 56 & Coverage and mean duration of each map (7 + 7) and transition fractions (42), from the winner-take-all sequence & Replication, hearing adjustment, PTA prediction, microstate classifiers \\
Whole-sequence microstate summaries & 2 & Mean segment duration, segment rate & Covariates (Table~\ref{tab:explain}) \\
Spectral baseline & 6 & Relative power in 5 bands and 1/$f$ slope of the EEG, mean over 15 common 10--20 channels & Replication baseline, covariates \\
Fast-mode fraction & 1 & Fraction of retained $|\lam|$ in $[0.70,0.90)$ & Primary Koopman feature: replication, hearing adjustment \\
$|\lam|$ quantiles & 21 & 20 quantiles of retained $|\lam|$ and the fraction retained & PTA prediction \\
Other Koopman candidates & 60 & $|\lam|$ bands and quantiles, mode frequencies, $|\lam|\times$frequency cells, mode composition, amplitude-weighted measures & Beyond the fast-mode fraction (Table~\ref{tab:S-beyond}) \\
Koopman classifier & -- & Full distribution of retained $|\lam|$, scored by Wasserstein distance to class prototypes & Classification \\
\bottomrule
\end{tabular}
\end{table}

\subsection*{Feature-level replication tests}
\label{sec:reptest}
The microstate and Koopman features (Table~\ref{tab:features}) were tested with the same procedure, with relative band power and 1/$f$ slope of the EEG spectrum acting as a baseline. The primary Koopman
feature was selected to be the \emph{fast-mode fraction}, the fraction of a subject's retained eigenvalues with $|\lam|\in[0.70,0.90)$ (decay time constants of about 14--47~ms). Each of these sets, and each Koopman feature family in the supplement, was treated as a separate family for multiple-comparison correction.

\emph{Per-dataset tests.} Within each dataset, tinnitus and control subjects were compared with a two-sided Mann--Whitney $U$ test. The effect size is the rank-biserial correlation
\begin{equation}
  \rrb = \frac{2U}{n_T n_C} - 1,
\end{equation}
where $n_T$ and $n_C$ are the group sizes; $\rrb>0$ means higher values in tinnitus, and $(1+\rrb)/2$ is the probability that a randomly chosen tinnitus subject exceeds a randomly chosen control. Its standard error was taken from the null variance, $\mathrm{SE}=\sqrt{(n_T+n_C+1)/(3n_Tn_C)}$, and each two-sided $p$-value was converted to a signed normal score $z_i=\operatorname{sign}(\rrb)\,\Phi^{-1}(1-p_i/2)$.

\emph{Pooling and correction.} The three datasets were combined by Stouffer's method,
$Z=\sum_{i=1}^{3} z_i/\sqrt{3}$, with equal weight per dataset so that the largest dataset (D1) cannot determine the result alone, and $Z$ was converted to a two-sided pooled $p$. Pooled $p$-values were corrected for the false discovery rate with the Benjamini--Hochberg procedure within each feature family (e.g., across the 56 microstate statistics), giving $q$. For reporting, the pooled effect is the inverse-variance weighted mean of the per-dataset effects, and between-dataset heterogeneity is summarized by $I^2=\max\{0,(Q-2)/Q\}$, with $Q$ Cochran's statistic. As a sensitivity analysis, the per-dataset effects were also pooled with a random-effects model, which adds the estimated between-dataset variance $\tau^2$ to each dataset's variance, so that the pooled effect, its interval and its $p$-value come from the same model; $q$ was then obtained as above and the same replication criterion applied (Table~\ref{tab:S-re}). With three datasets $\tau^2$ is poorly determined, so this model was not used as the primary analysis.

\emph{Replication criterion.} A feature was said to \emph{replicate} if its effect (tinnitus - control) had the same sign in all three datasets and its pooled $q<0.05$. Requiring the same sign excludes features whose pooled significance is carried by one dataset; under the null hypothesis, the same sign occurs by chance for 25\% of features. Within-dataset significance (Figs.~\ref{fig:S-box} and~\ref{fig:S-msdyn}) was assessed separately, with Benjamini--Hochberg correction across the 56 statistics of each dataset.

\emph{Hearing adjustment.} To test whether a group difference is explained by hearing level, each feature's ranks were regressed by ordinary least squares on group, PTA and, where available, age and sex (D1: all three; D2: age in three categories; D3: PTA only), with covariates z-scored within each dataset. The adjusted effect is $2\hat\beta_{\text{group}}/N$, which equals $\rrb$ without covariates, with standard error $2\,\mathrm{SE}(\hat\beta_{\text{group}})/N$ and $p$ from the $t$-test of $\hat\beta_{\text{group}}$; subjects missing a covariate were excluded from that model. Pooling, correction and the replication criterion were then applied exactly as above. The same model, with additional covariates, was used to test whether the fast-mode effect is explained by microstate or spectral measures (Table~\ref{tab:explain}).

\emph{Choice of the fast band.} The band $[0.70,0.90)$ was identified from the group differences in the quantile functions of $|\lam|$ (Fig.~\ref{fig:lambda}) on all three datasets, so its replication is not by itself an independent test. We therefore also chose the band without the dataset it is tested on. For each dataset, the band was selected on the other two (same sign in both, largest pooled unadjusted Stouffer $|z|$ among bands $[0.70,h)$, $h$ on a 0.01 grid) and its fraction tested on the held-out dataset. The three held-out tests use disjoint data and were combined by Stouffer's method, one-sided in the direction chosen in training.

\emph{Surrogate control.} To test whether the fast-mode fraction depends on structure beyond the second-order statistics of the activations, multivariate phase-randomized surrogates \cite{prichard_generating_1994} of each subject's 60~s activation window $\mathbf{a}(t)$ were generated. The Fourier transform of each of the 7 activations was multiplied by $e^{i\phi(f)}$ with one random phase $\phi(f)$ per frequency, shared across the 7 channels, and inverted. Such surrogates preserve every auto- and cross-spectrum of $\mathbf{a}(t)$ (hence all linear correlations within and between maps at every lag) and destroy all higher-order temporal structure. The Koopman operator was refit on each surrogate with the same observables, embedding and least-squares fit, and the fast-mode fraction recomputed; five surrogates per subject were averaged. The group tests of the replication analysis (\emph{Pooling and correction}) were applied to the original fraction, the surrogate fraction and their difference (Table~\ref{tab:S-surr}).

\subsection*{Classification}
\label{sec:clf}
\subsubsection*{Microstates}
The 56 microstate statistics (Eqs.~\eqref{eq:coverage}--\eqref{eq:transition}) were used as features for classification. We trained a Linear Support Vector Machine (SVM) on these features, with class-balanced weights and default hyperparameters. We also trained an RBF SVM, $\ell_2$ logistic regression, and a random forest on the same features, all with class-balanced weights and default hyperparameters.

\subsubsection*{Koopman}
We used the 1-D Wasserstein distance to classify subjects based on their Koopman eigenvalue distributions. Each class was represented by the pooled $|\lam|$ distribution of its training subjects, and a test subject was scored by the difference of its 1-D Wasserstein distances to the control and tinnitus prototypes. 

\subsubsection*{Protocols} 
We report 5-fold cross-validation results within each dataset, and training on one dataset and testing on another in all six directions, without retraining or fine-tuning. Due to the class imbalanced nature of all three datasets, we report AUROC, AUPRC, MCC and balanced accuracy, with 95\% class-stratified bootstrap CIs (1000 resamples) and one-sided label-permutation $p$ for AUROC.

\subsection*{Hearing-level analyses}
\label{sec:hearing}
PTA was the four-frequency average (0.5, 1, 2 and 4~kHz) where thresholds were available, otherwise the PTA or mean hearing level reported with the dataset, averaged over both ears. We used PTA alone as a classifier to identify the contribution of hearing level to group separation. We then showed classification performance on 1:1 PTA-matched subsets (caliper 3~dB), analyzed per dataset and pooled across datasets. To test whether a feature set carries a hearing signature, PTA was predicted from it by ridge regression (standardized features; regularization chosen within each training fold by efficient leave-one-out cross-validation) with 5-fold out-of-fold predictions; we report the Spearman $\rho$ between predicted and true PTA and a permutation $p$ (200 permutations of PTA, whole pipeline refit). Classifier scores were correlated with tinnitus severity (Table~\ref{tab:datasets}) by Spearman $\rho$ within tinnitus subjects, with Benjamini--Hochberg correction.

% =====================================================================

\section*{Data availability}
All three datasets are publicly available. D1 is available via the link provided in the original publication~\cite{wang_cross-subject_2023}; D2 is available from Mendeley Data at \url{https://doi.org/10.17632/kj443jc4yc.2}~\cite{cuevas-romero_electroencephalography-based_2022}; D3 is available from Mendeley Data at \url{https://doi.org/10.17632/fj7sskjdt7.5}~\cite{torres-torres_characterization_2023}. The derived per-subject features (microstate statistics, spectral features and Koopman eigenvalues) underlying the figures and tables are available from the corresponding author on reasonable request.

\section*{Code availability}
The underlying code for this study is not publicly available but may be made available to qualified researchers on reasonable request from the corresponding author.

\section*{Acknowledgements}
This work was not supported by any external funding.

\section*{Author contributions}
A.B. and G.C., designed the study and identified the effect and required embeddings. A.B. and performed the analysis. A.U., Y.X and M.L. contributed to the preprocessing, organization and review of the analysis. A.M contributed to the clinical relevance of the results. All authors read and approved the final manuscript.

\section*{Competing interests}
All authors declare no financial or non-financial competing interests.

\linespread{1.0}\selectfont
\bibliographystyle{unsrt}
\bibliography{references}

\clearpage
\section*{Supplementary Information}
\renewcommand{\thefigure}{S\arabic{figure}}
\setcounter{figure}{0}
\renewcommand{\thetable}{S\arabic{table}}
\setcounter{table}{0}

\begin{figure}[!htbp]
\centering
% \figfile{fig2c_selffit_vs_custo_maps}{0.85\columnwidth}
\includegraphics[width=0.85\columnwidth]{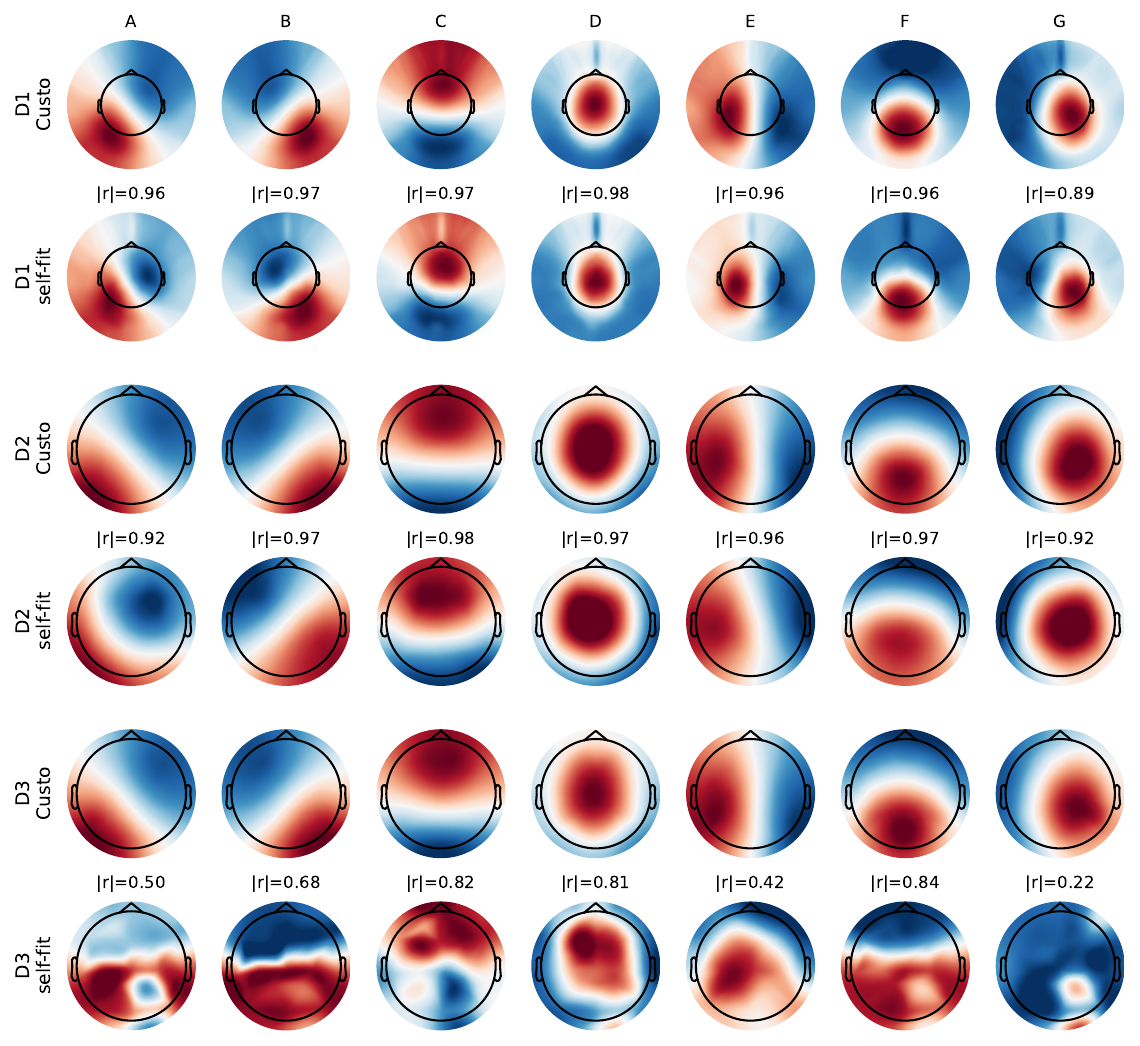}
\caption{$k=7$ maps fit to each whole dataset (self-fit), relabeled by matching to the Custo et al.
template \cite{custo_electroencephalographic_2017}, next to the template on the same montage; titles give $|r|$. D1 and D2 closely reproduce the
template (mean $|r|=0.96$); D3 does not (0.61).}
\label{fig:S-selffit}
\end{figure}

\begin{figure}[!htbp]
\centering
% \figfile{fig2b_group_selffit_maps}{0.85\columnwidth}
\includegraphics[width=0.85\columnwidth]{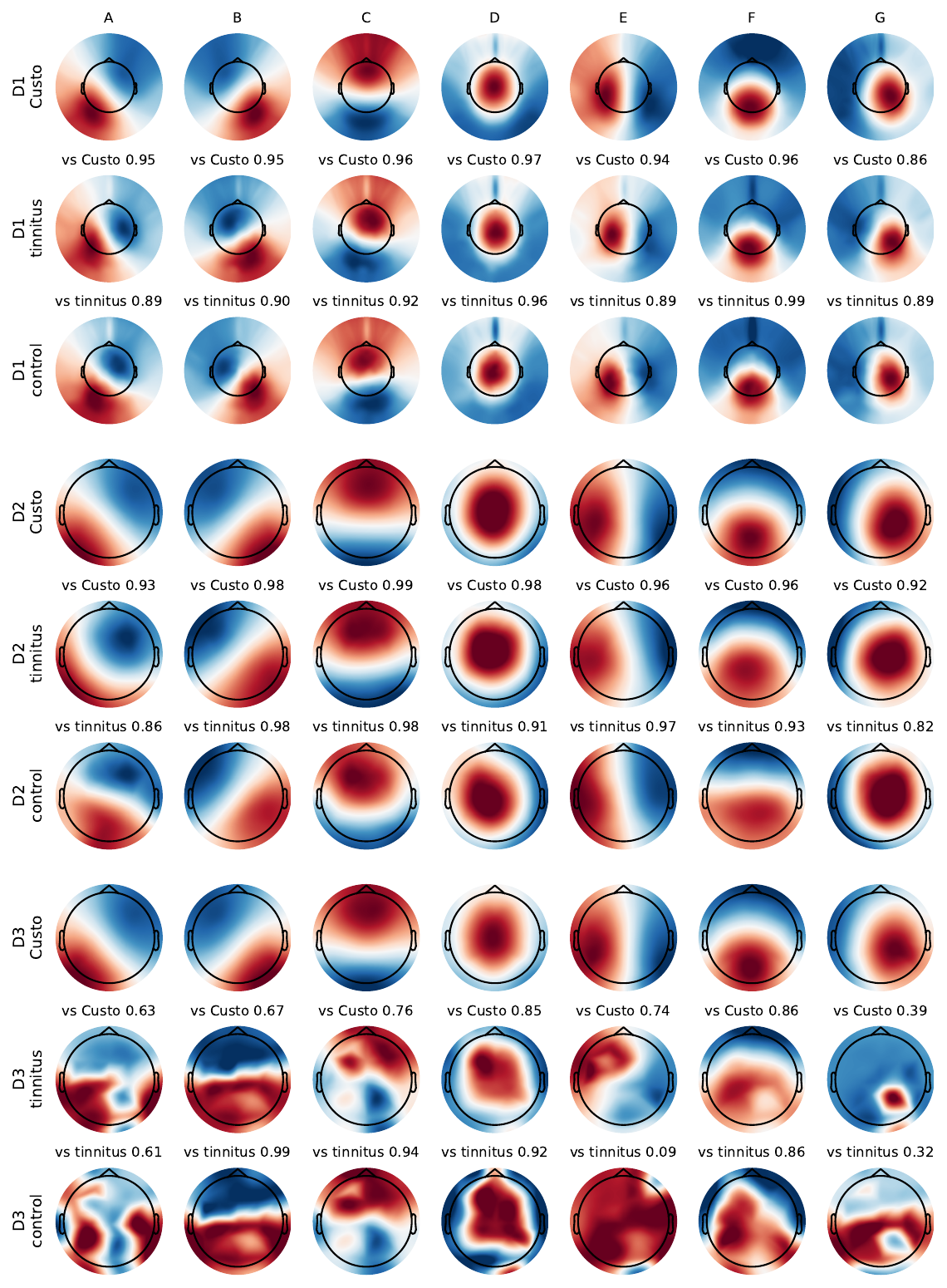}
\caption{$k=7$ maps fit separately to tinnitus and control subjects of each dataset, labeled by
matching to the template.}
\label{fig:S-groupmaps}
\end{figure}

\begin{figure*}[!htbp]
\centering
% \figfile{figS_noisy_channels}{0.95\textwidth}
\includegraphics[width=0.95\textwidth]{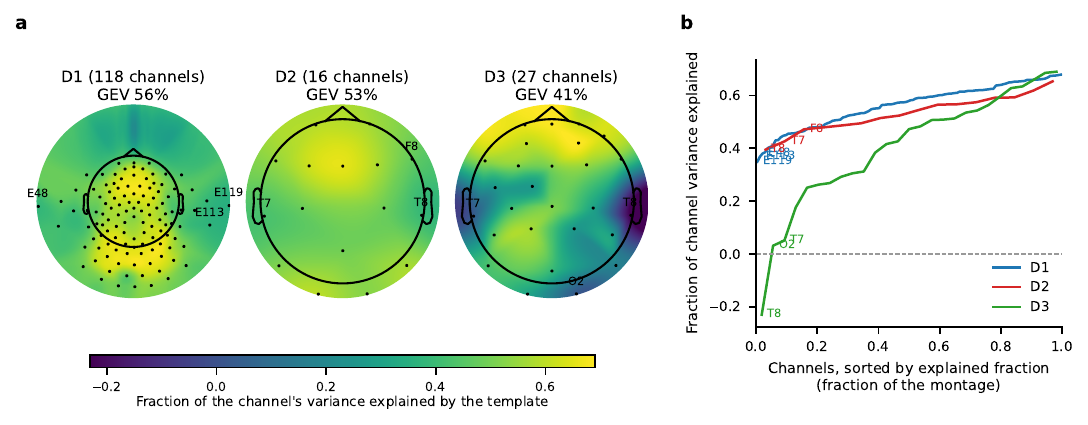}
\caption{Channels that the template does not explain. (a) For each dataset, the fraction of each channel's variance that the winner-take-all back-fit of the seven template maps explains, pooled over subjects (channels are standardized, so each contributes equally to the total variance); the three least-explained channels are labeled. (b) The same fractions sorted, against the channel's rank as a fraction of the montage. The temporal channels are explained worst on all three montages. In D3 the temporal channels T8 and T7 and the occipital O2 are essentially unexplained (explained fractions $-0.23$, 0.05 and 0.03, against a minimum of 0.35 in D1 and 0.40 in D2); excluding these three channels raises D3's GEV from 41\% to 47\%, and the median channel of D3 (0.47) remains explained less well than in D1 (0.58) or D2 (0.53).}
\label{fig:S-noisychannels}
\end{figure*}

\begin{figure}[!htbp]
\centering
% \figfile{figS2_decay_band_across_embeddings}{0.8\columnwidth}
\includegraphics[width=0.8\columnwidth]{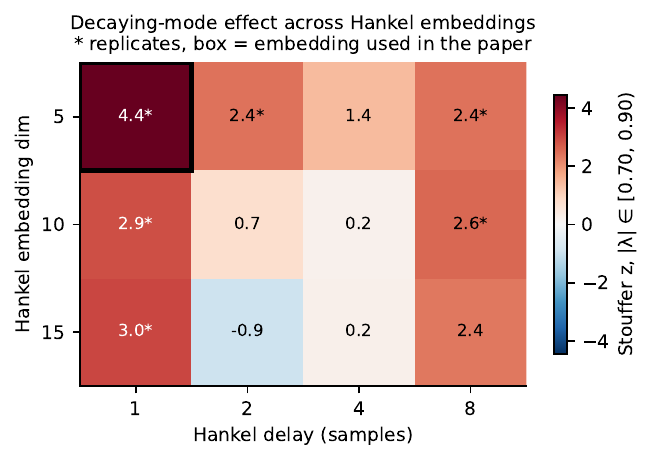}
\caption{Fast-mode fraction effect (Stouffer $z$) for every Hankel embedding dimension and delay. *:
replicates; box: embedding used in the paper.}
\label{fig:S-robust}
\end{figure}

\begin{figure}[!htbp]
\centering
% \figfile{figS4_threshold_calibration_roc}{\columnwidth}
\includegraphics[width=\columnwidth]{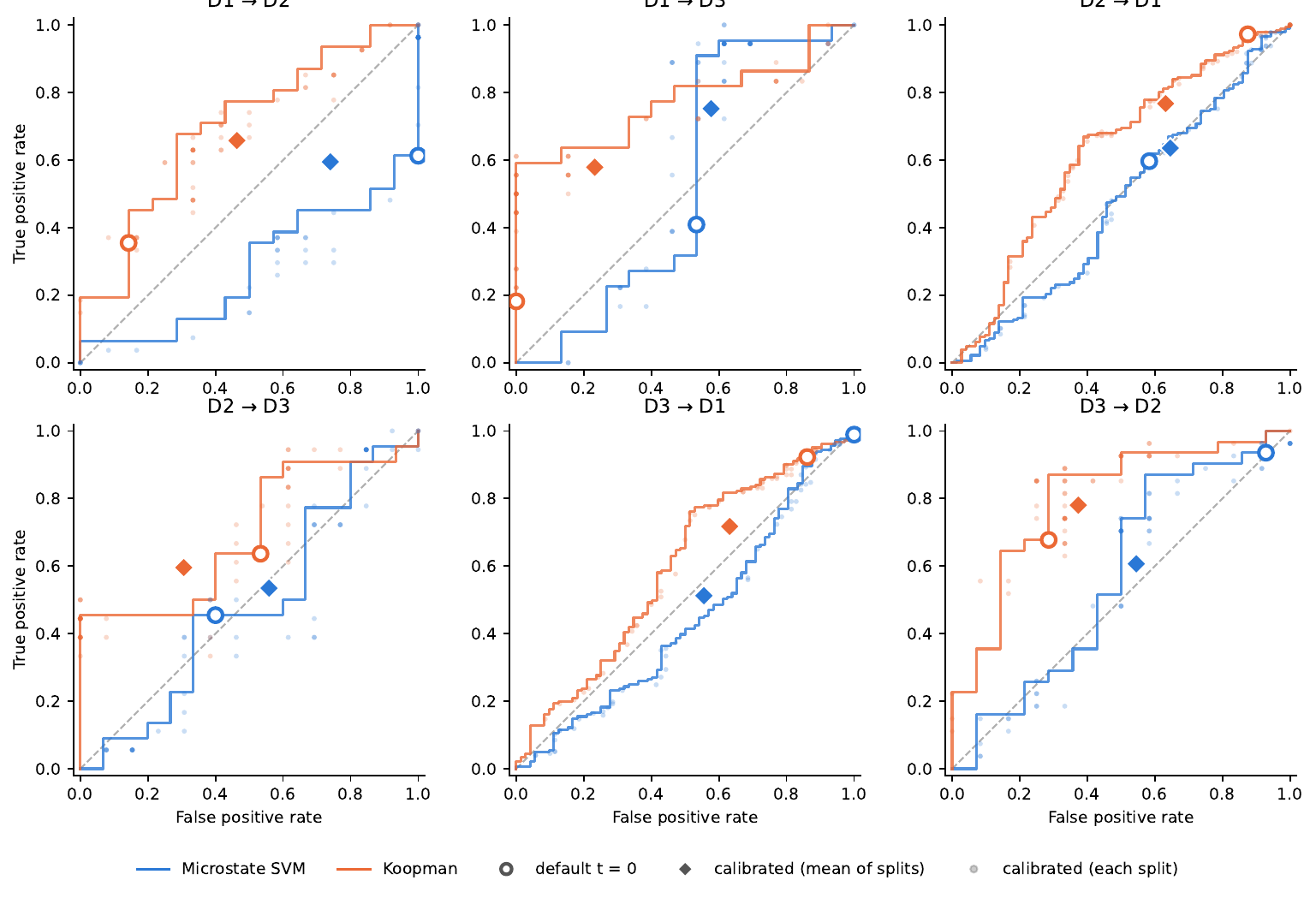}
\caption{Decision-threshold calibration on three labeled test-site subjects per class.}
\label{fig:S-calib}
\end{figure}

\begin{figure}[!htbp]
\centering
% \figfile{figS1a_config_worst_mcc_distribution}{0.8\columnwidth}
\includegraphics[width=0.8\columnwidth]{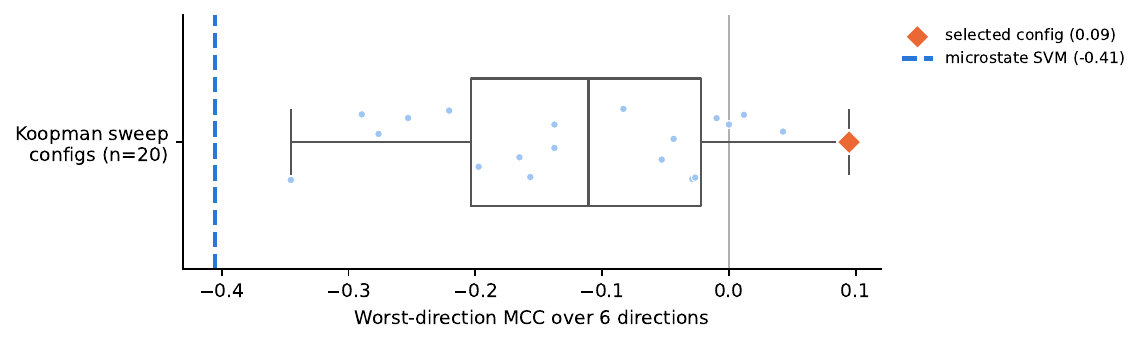}
\caption{Worst-direction MCC of all configurations in the hyperparameter sweep; the configuration used
is marked.}
\label{fig:S-sweep}
\end{figure}

\begin{figure}[!htbp]
\centering
% \figfile{fig3b_effect_size_matrices}{\columnwidth}
\includegraphics[width=\columnwidth]{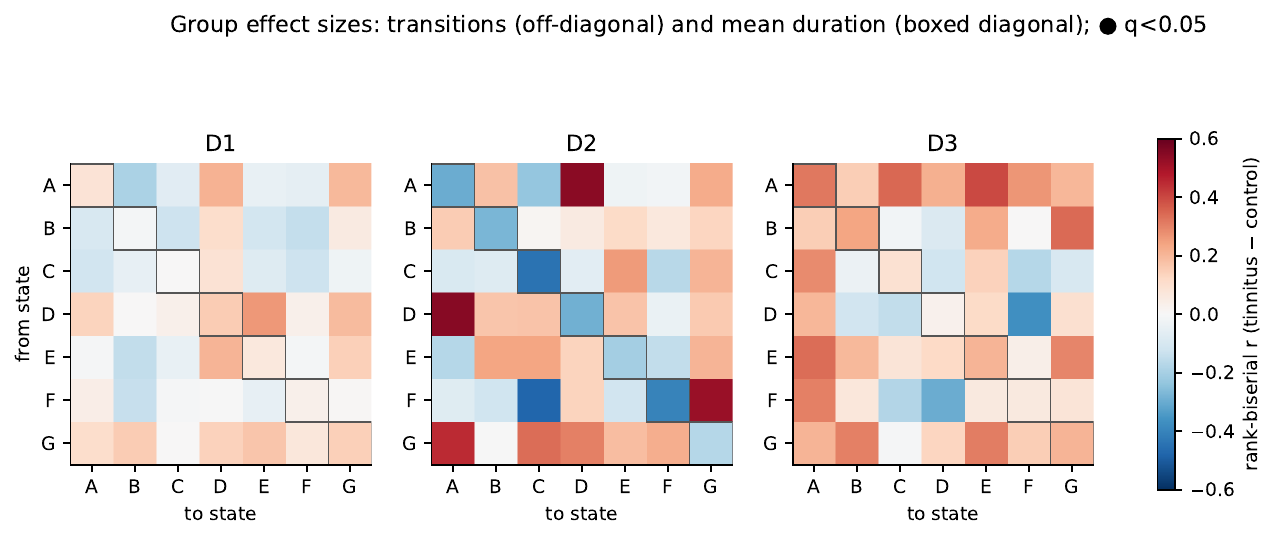}
\caption{Microstate group effects per dataset. Off-diagonal: $\rrb$ of transition probability
$P(i\to j)$; diagonal: $\rrb$ of mean duration. Positive (red): higher in tinnitus. No effect survives
FDR correction within any dataset.}
\label{fig:S-msdyn}
\end{figure}

\begin{figure}[!htbp]
\centering
% \figfile{fig4_cross_dataset_roc}{\columnwidth}
\includegraphics[width=\columnwidth]{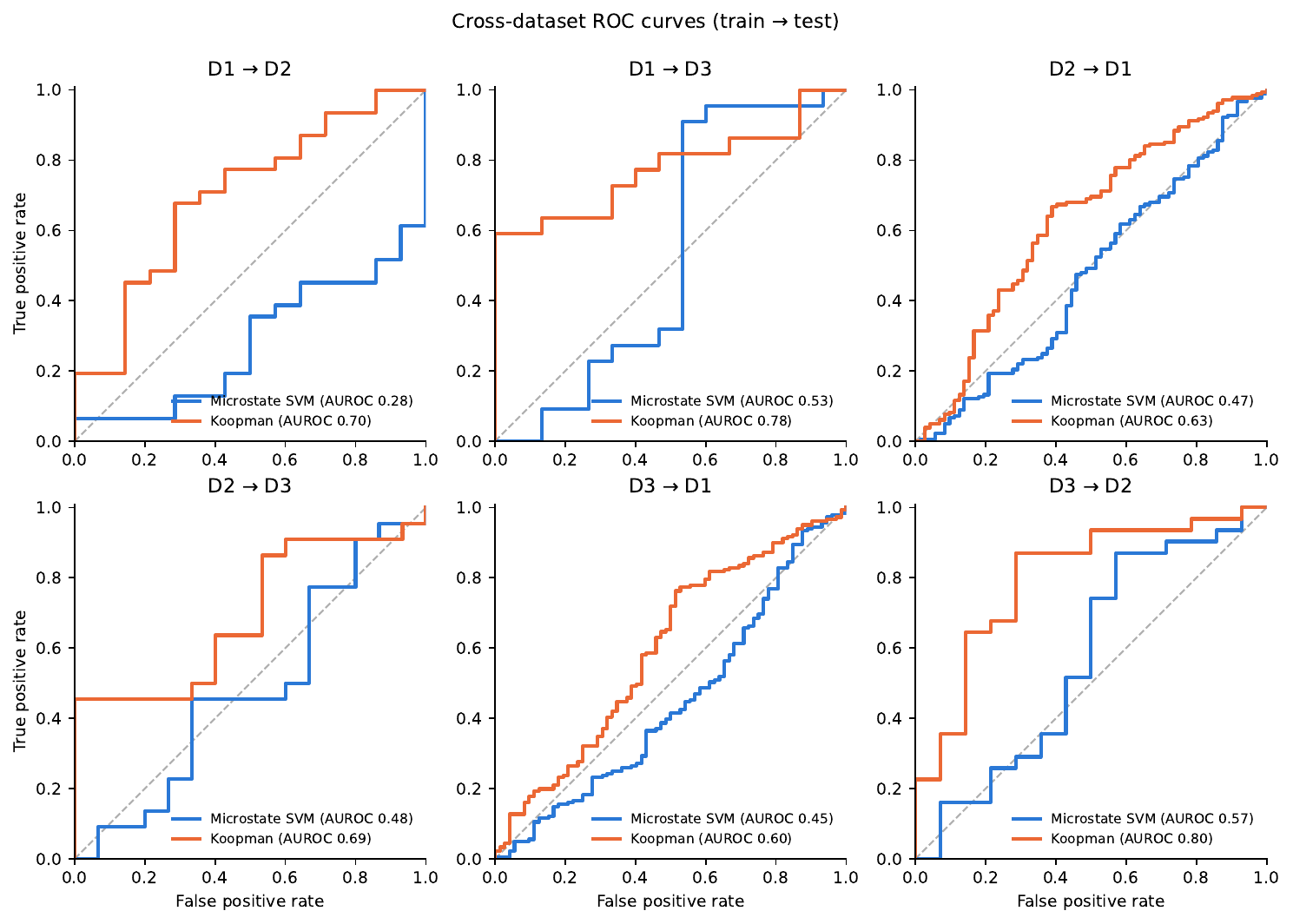}
\caption{Cross-dataset ROC curves, Koopman vs.\ linear-SVM microstate baseline, all six train $\to$
test directions.}
\label{fig:S-roc}
\end{figure}

\begin{figure}[!htbp]
\centering
% \figfile{fig6_matched_auroc_robustness}{\columnwidth}
\includegraphics[width=\columnwidth]{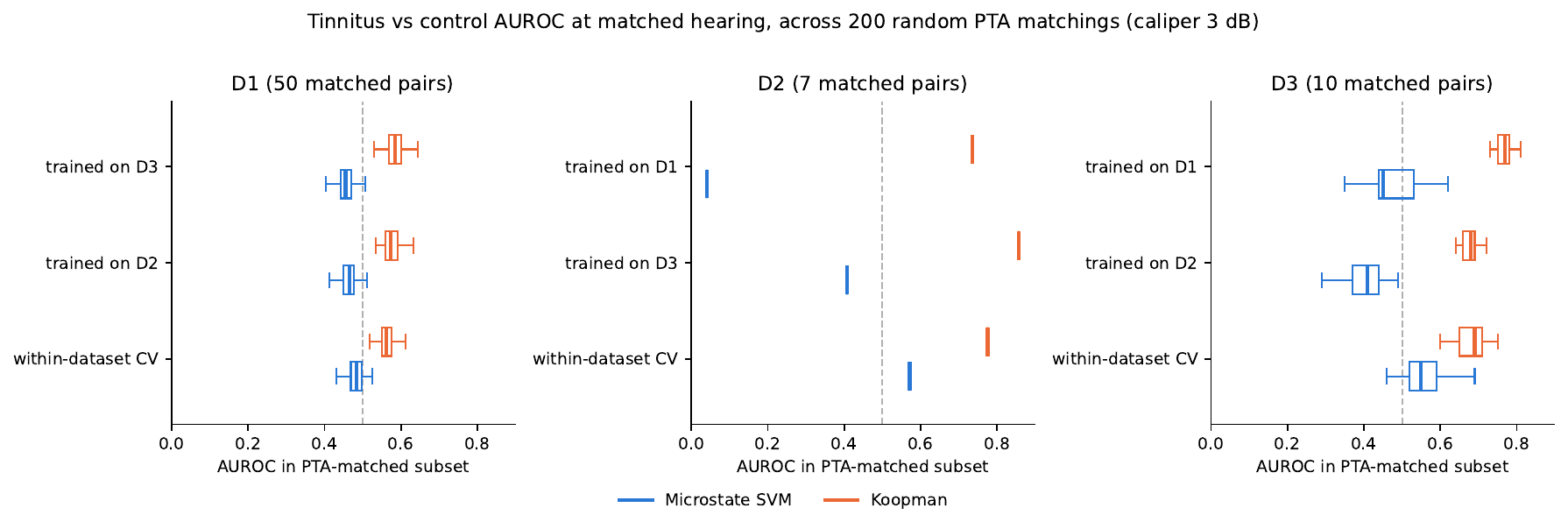}
\caption{AUROC in PTA-matched subsets across 200 random matchings, per dataset and score source.}
\label{fig:S-matched}
\end{figure}

\begin{figure}[!htbp]
\centering
% \figfile{fig8_feature_replication}{\columnwidth}
\includegraphics[width=\columnwidth]{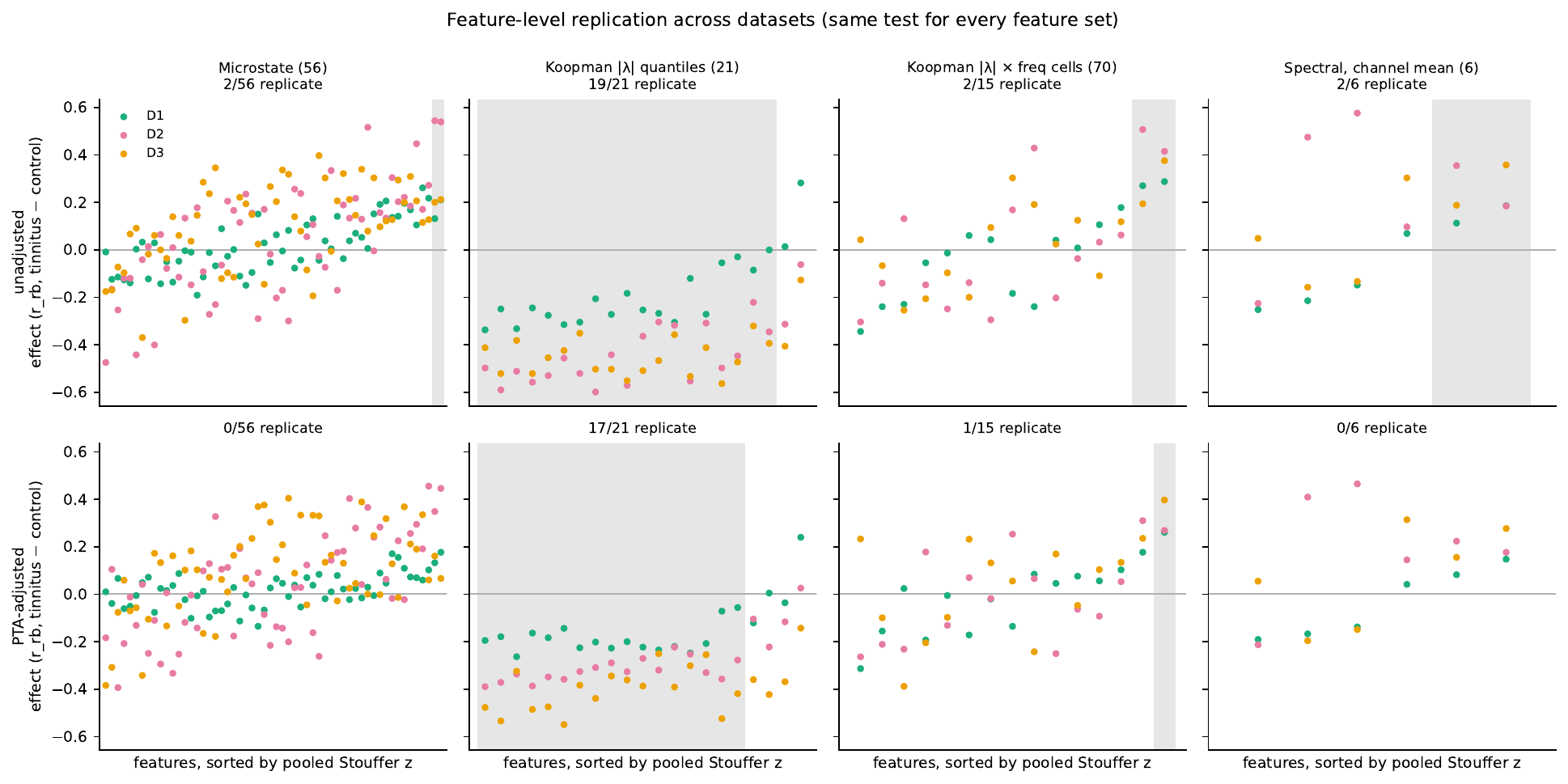}
\caption{Feature-level replication for every feature, one procedure for every feature set. Each column
is a feature, sorted by pooled Stouffer $z$; dots: per-dataset $\rrb$; gray: replicating features. Top:
unadjusted; bottom: adjusted for PTA (+ age, sex).}
\label{fig:S-rep}
\end{figure}

\begin{figure}[!htbp]
\centering
% \figfile{fig3a_coverage_duration_by_group}{\columnwidth}
\includegraphics[width=\columnwidth]{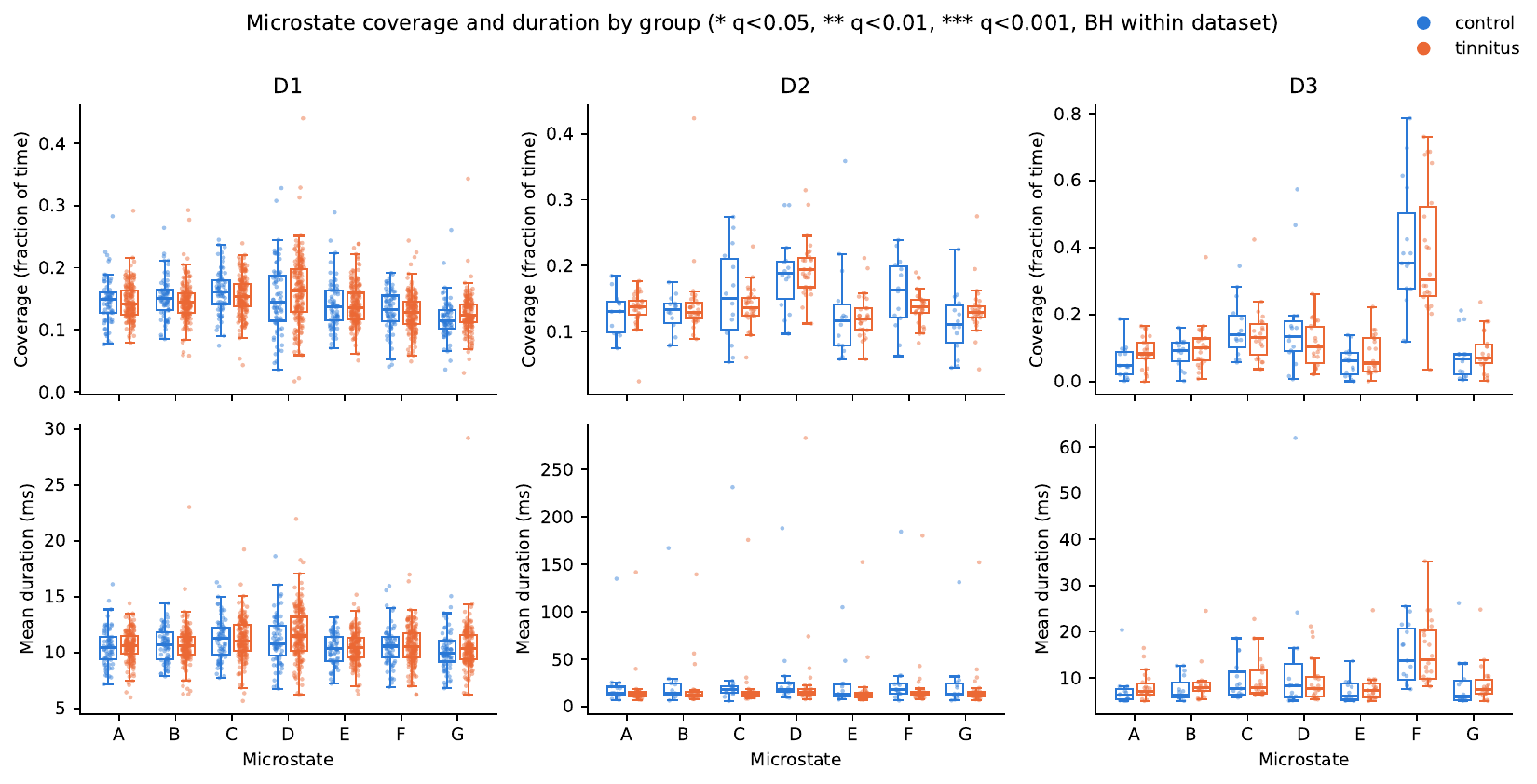}
\caption{Coverage and mean duration of each microstate by group and dataset. Stars: FDR $q<0.05$
within dataset (none).}
\label{fig:S-box}
\end{figure}

\begin{figure}[!htbp]
\centering
% \figfile{fig10_feature_vs_pta}{\columnwidth}
\includegraphics[width=\columnwidth]{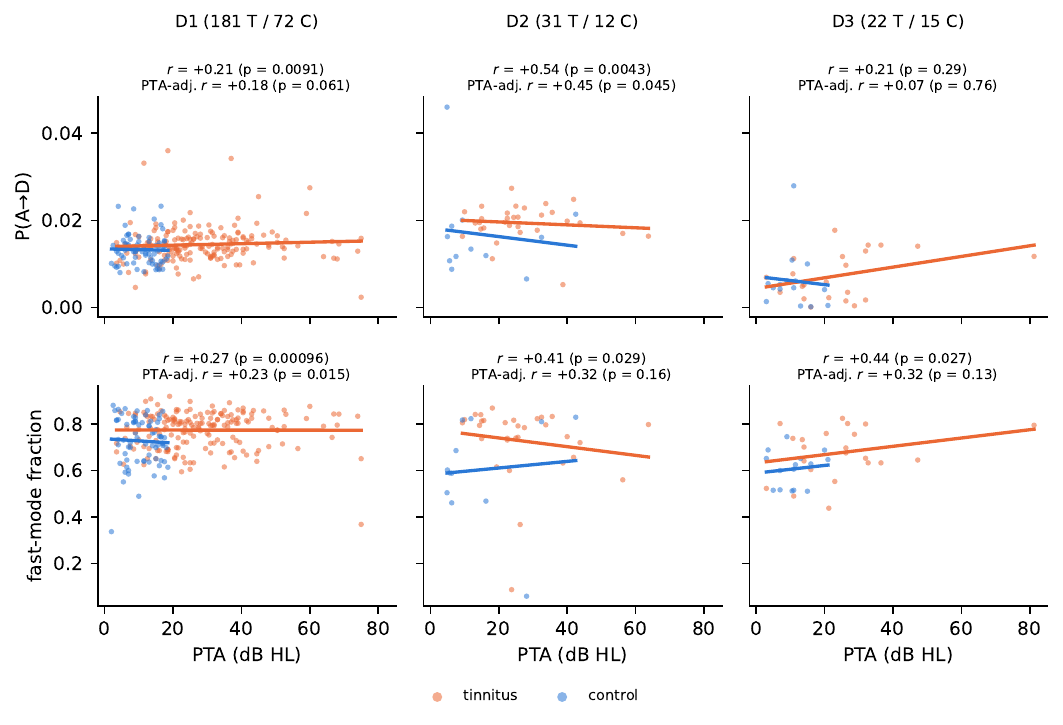}
\caption{Raw data behind Fig.~\ref{fig:pta}: $P(\mathrm{A}\to\mathrm{D})$ (top) and the fast-mode fraction
(bottom) against PTA, per dataset (subjects with PTA; T / C: tinnitus / control), with within-group
least-squares lines. Above each panel: the dataset's unadjusted and PTA-adjusted group effects.}
\label{fig:S-ptascatter}
\end{figure}

\begin{figure*}[!tbp]
\centering
% \figfile{fig7b_lambda_frequency_effects}{0.95\textwidth}
\includegraphics[width=0.95\textwidth]{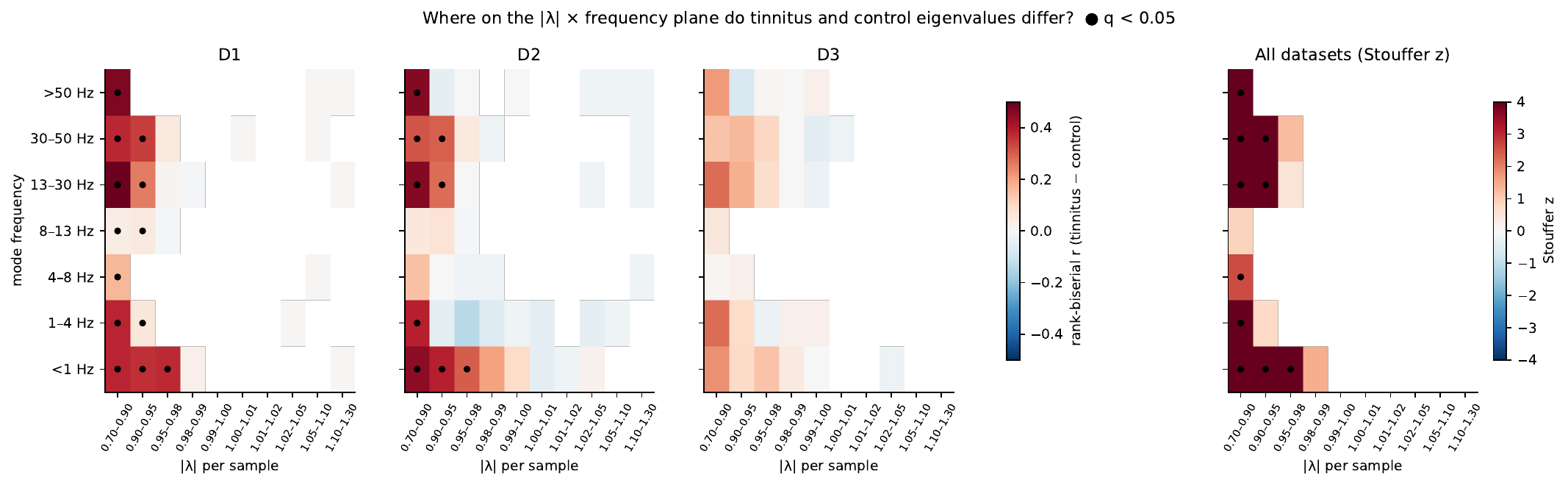}
\caption{Group effect on the fraction of retained eigenvalues in each $|\lam|\times$mode-frequency cell,
per dataset and pooled (Stouffer $z$). Dots: $q<0.05$. Mode frequencies are not EEG bands: fast modes are
mostly quadratic and 7--23~Hz wide (main-text Fig.~\ref{fig:composition}).}
\label{fig:S-lamfreq}
\end{figure*}

\begin{table}[!htbp]
\caption{Fast band chosen on two datasets, tested on the third.}
\label{tab:S-lodo}
\centering
\footnotesize
\setlength{\tabcolsep}{3pt}
\begin{tabular}{lcccc}
\toprule
Held out & Band & Train $z$ & $r$ & $r$, PTA-adj.\ ($p$) \\
\midrule
D1 & $[0.70, 0.81)$ & 3.60 & 0.36 & 0.27 (0.002) \\
D2 & $[0.70, 0.81)$ & 4.97 & 0.48 & 0.28 (0.10) \\
D3 & $[0.70, 0.84)$ & 4.98 & 0.45 & 0.36 (0.047) \\
\midrule
Combined & & & $z=5.36$ & $z=3.40$ (0.0003) \\
\bottomrule
\end{tabular}
\\[2pt]\raggedright\footnotesize Band: chosen on the other two datasets, $[0.70,h)$ with the largest pooled unadjusted Stouffer $|z|$ (same sign in
both training datasets), $h$ on a 0.01 grid. $r$: held-out rank-biserial effect; $p$: one-sided, in the direction chosen in training. Combined:
Stouffer $z$ over the three disjoint held-out tests. Letting both band edges vary gave a combined PTA-adjusted
$z=2.84$ ($p=0.002$). Paper band $[0.70,0.90)$, chosen on all three: PTA-adjusted $r=$ 0.23 / 0.32 / 0.32.
\end{table}

\begin{table}[!htbp]
\caption{Sensitivity to regularizing the Koopman fit (truncated SVD).}
\label{tab:S-trunc}
\centering
\footnotesize
\setlength{\tabcolsep}{2.5pt}
\begin{tabular}{lccccc}
\toprule
 & Untrunc. & $10^{-6}$ & $10^{-5}$ & $10^{-4}$ & $10^{-3}$ \\
\midrule
Median rank kept                & 175 & \prov{151} & \prov{126} & \prov{98} & \prov{58} \\
$\rho$ fast-mode fraction vs.\ untrunc. & 1 & \prov{0.91} & \prov{0.89} & \prov{0.75} & \prov{0.58} \\
Max.\ change under perturbation & \prov{0.16} & \prov{0.02} & \prov{0} & \prov{0} & \prov{0} \\
Fast-mode $r$, PTA-adj.         & \prov{0.26} & \prov{0.25} & \prov{0.26} & \prov{0.26} & \prov{0.29} \\
\quad $p$                       & \prov{0.002} & \prov{0.005} & \prov{0.003} & \prov{0.002} & \prov{$<$0.001} \\
Quantiles replicating, PTA-adj. & \prov{17/21} & \prov{16/21} & \prov{16/21} & \prov{11/21} & \prov{12/21} \\
Mean AUROC (cross-dataset)      & \prov{0.70} & \prov{0.70} & \prov{0.68} & \prov{0.65} & \prov{0.67} \\
Worst AUROC                     & \prov{0.60} & \prov{0.59} & \prov{0.55} & \prov{0.55} & \prov{0.52} \\
\bottomrule
\end{tabular}
\\[2pt]\raggedright\footnotesize Columns: relative singular-value threshold. $\rho$: minimum over datasets.
Perturbation: $10^{-6}$ of the signal SD; largest change of any subject's fast-mode fraction.
\end{table}

\emph{Regularized fit.} The pseudo-inverse in \eqref{eq:koopman} keeps every singular value of $\boldsymbol\Psi$, whose condition number is large (median $\approx10^{7}$--$10^{8}$; up to $10^{16}$ for one D1 subject). Every subject's operator, native and on every cross-dataset test side, was refit with a truncated SVD keeping singular values above $10^{-6}$--$10^{-3}$ of the largest, using the projected form $\mathbf{U}_r^\top\boldsymbol\Psi_+\mathbf{V}_r\mathbf{S}_r^{-1}$, whose eigenvalues are the nonzero eigenvalues of the rank-$r$ fit. Per-subject stability was assessed by refitting after perturbing the activations by $10^{-6}$ of their standard deviation. The hearing-adjusted fast-mode effect was \prov{0.25--0.29} at every threshold and truncation removed the per-subject instability (Table~\ref{tab:S-trunc}). The broader shift of the $|\lam|$ quantiles was less robust (adjusted replication \prov{17/21} untruncated, \prov{12/21} at the strongest truncation), which is why the fast-mode fraction is the primary feature.

\begin{table}[!htbp]
\caption{Koopman features beyond the fast-mode fraction: number replicating.}
\label{tab:S-beyond}
\centering
\footnotesize
\setlength{\tabcolsep}{2.5pt}
\begin{tabular}{lcccc}
\toprule
Family (testable) & Unadj. & PTA-adj. & + fast-mode fr. & $|\rho|$ \\
\midrule
$|\lam|$ quantiles (21)                & \prov{19} & \prov{17} & \prov{0} & \prov{0.74} \\
Other $|\lam|$ bands (4)               & \prov{2}  & \prov{2}  & \prov{0} & \prov{0.60} \\
Mode-frequency distribution (8)        & \prov{3}  & \prov{2}  & \prov{0} & \prov{0.45} \\
$|\lam|\times$mode-freq.\ cells (15)  & \prov{2}  & \prov{1}  & \prov{0} & \prov{0.19} \\
Fast-mode composition (10)             & \prov{2}  & \prov{2}  & \prov{0} & \prov{0.13} \\
Amplitude-weighted (2)                 & \prov{2}  & \prov{0}  & \prov{0} & \prov{0.57} \\
\bottomrule
\end{tabular}
\\[2pt]\raggedright\footnotesize Columns: replicating features without adjustment, after PTA (+ age, sex),
and after PTA (+ age, sex) and the fast-mode fraction. $|\rho|$: median absolute Spearman correlation with the
fast-mode fraction.
\end{table}

\begin{table}[!htbp]
\caption{All microstate classifiers (56 statistics), AUROC / MCC.}
\label{tab:S-clf}
\centering
\footnotesize
\setlength{\tabcolsep}{3pt}
\begin{tabular}{lcccc}
\toprule
 & Linear SVM & Logistic reg. & RBF SVM & Random forest \\
\midrule
D1 (CV)     & 0.57 / 0.06    & 0.60 / 0.08    & 0.57 / 0.07    & 0.58 / 0.00 \\
D2 (CV)     & 0.64 / 0.27    & 0.63 / 0.34    & 0.74 / 0.24    & 0.78 / 0.37 \\
D3 (CV)     & 0.54 / 0.22    & 0.52 / 0.06    & 0.57 / $-0.01$ & 0.56 / $-0.03$ \\
\midrule
D1 $\to$ D2 & 0.28 / $-0.41$ & 0.32 / $-0.22$ & 0.62 / 0.21    & 0.66 / 0.22 \\
D1 $\to$ D3 & 0.53 / $-0.12$ & 0.55 / $-0.10$ & 0.47 / 0.00    & 0.66 / 0.30 \\
D2 $\to$ D1 & 0.47 / 0.01    & 0.47 / $-0.02$ & 0.60 / 0.10    & 0.61 / 0.09 \\
D2 $\to$ D3 & 0.48 / 0.05    & 0.54 / 0.17    & 0.74 / 0.25    & 0.60 / 0.08 \\
D3 $\to$ D1 & 0.45 / $-0.06$ & 0.45 / 0.00    & 0.57 / $-0.04$ & 0.50 / 0.00 \\
D3 $\to$ D2 & 0.57 / 0.01    & 0.54 / 0.09    & 0.78 / 0.32    & 0.61 / $-0.10$ \\
\midrule
Cross-dataset mean & 0.46 / $-0.08$ & 0.48 / $-0.01$ & 0.63 / 0.14 & 0.61 / 0.10 \\
\bottomrule
\end{tabular}
\\[2pt]\raggedright\footnotesize Class-balanced weights, default hyperparameters; MCC at the default threshold. CV: within-dataset 5-fold cross-validation.
\end{table}

\begin{table}[!htbp]
\caption{Eigenvalue magnitude vs.\ phase vs.\ the full complex cloud.}
\label{tab:S-cloud}
\centering
\footnotesize
\setlength{\tabcolsep}{3pt}
\begin{tabular}{lccc}
\toprule
 & $|\lam|$ (paper) & $\arg\lam$ & Full cloud \\
\midrule
Cross-dataset AUROC, mean       & \prov{0.70} & \prov{0.57} & \prov{0.63} \\
Cross-dataset AUROC, worst      & \prov{0.60} & \prov{0.48} & \prov{0.51} \\
Cross-dataset MCC, mean         & \prov{0.21} & \prov{0.05} & \prov{0.05} \\
Within-dataset AUROC, D1       & \prov{0.61} & \prov{0.64} & \prov{0.65} \\
Within-dataset AUROC, D2       & \prov{0.76} & \prov{0.64} & \prov{0.69} \\
Within-dataset AUROC, D3       & \prov{0.73} & \prov{0.54} & \prov{0.56} \\
Site / group shift              & \prov{1.4} & \prov{2.3} & \prov{2.2} \\
\bottomrule
\end{tabular}
\\[2pt]\raggedright\footnotesize Same Wasserstein-prototype classifier and $|\lam|$ filter; full cloud:
sliced Wasserstein on (Re, Im). Site / group shift: mean Wasserstein distance between datasets (same group)
divided by that between groups (same dataset), on pooled kept eigenvalues.
\end{table}

\begin{figure*}[!tbp]
\centering
% \figfile{figS_cloud_density_difference}{\textwidth}
\includegraphics[width=\textwidth]{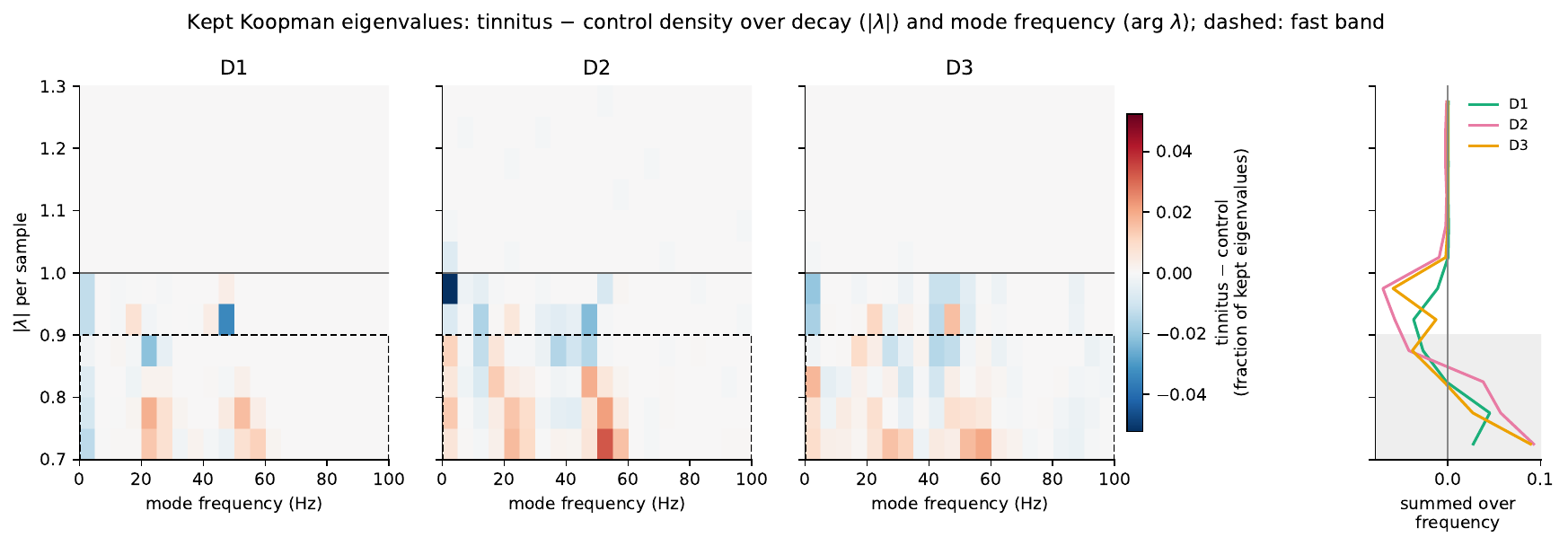}
\caption{Tinnitus minus control density of kept eigenvalues over decay ($|\lam|$) and mode frequency
($\arg\lam$), per dataset (dashed: fast band), and summed over frequency (right). The excess of
fast-decaying eigenvalues is shared across datasets; its distribution over mode frequency is not.}
\label{fig:S-cloud}
\end{figure*}

\begin{figure}[!htbp]
\centering
% \figfile{figS_linear_share_vs_frequency}{0.8\columnwidth}
\includegraphics[width=0.8\columnwidth]{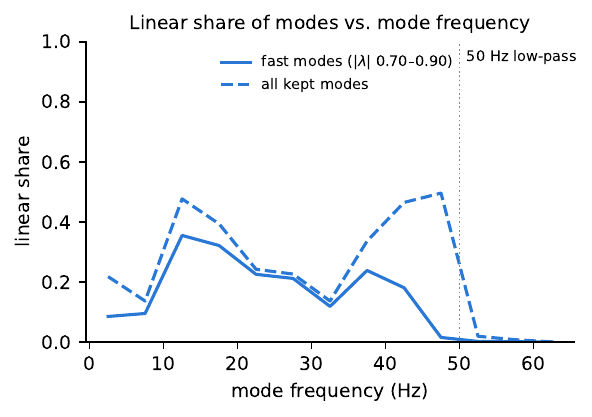}
\caption{Linear share of Koopman modes by mode frequency, for fast modes ($|\lam|\in[0.70,0.90)$) and all
retained modes (eigenvectors of the operator on unit-variance observables). Only the linear part of a mode
oscillates at the frequency of the scalp field; the linear share of fast modes stays below 0.36 at every mode
frequency and is near zero above 45~Hz, and their spectral width is 7--23~Hz, so mode frequencies are not EEG
bands.}
\label{fig:S-linfreq}
\end{figure}

\begin{table}[!htbp]
\caption{Replication under Stouffer and random-effects pooling.}
\label{tab:S-re}
\centering
\footnotesize
\setlength{\tabcolsep}{3pt}
\begin{tabular}{lcccc}
\toprule
 & \multicolumn{2}{c}{Unadjusted} & \multicolumn{2}{c}{PTA-adjusted} \\
Feature set & Stouffer & Random eff. & Stouffer & Random eff. \\
\midrule
Microstate statistics (56) & 2 & 1 & 0 & 0 \\
Spectral, channel mean (6) & 2 & 1 & 0 & 0 \\
$|\lam|$ quantiles (21) & 19 & 15 & 17 & 15 \\
$|\lam|\times$mode-frequency cells (15) & 2 & 3 & 1 & 1 \\
Fast-mode fraction (1) & yes & yes & yes & yes \\
Beyond the fast-mode fraction (60) & 30 & 26 & 24 & 20 \\
\quad + adjusted for fast-mode fraction & & & 0 & 0 \\
\bottomrule
\end{tabular}
\\[2pt]\raggedright\footnotesize Number of features that replicate (same sign in all three datasets, pooled BH $q<0.05$ within the set). Random effects: DerSimonian--Laird, pooled $p$ from the random-effects estimate and its standard error. Microstate statistics replicating, unadjusted: Stouffer, $P(\mathrm{A}\to\mathrm{D})$ and $P(\mathrm{D}\to\mathrm{A})$; random effects, $P(\mathrm{D}\to\mathrm{E})$ (random-effects $q=0.051$, 0.20 and 0.049; $\tau^2=0.007$, 0.024 and 0). Spectral, unadjusted: Stouffer, relative $\alpha$ and $\beta$; random effects, relative $\alpha$. Fast-mode fraction, random effects: $r=0.31$ (95\% CI 0.17--0.44), $p=9\times10^{-6}$ unadjusted; $r=0.26$ (0.10--0.41), $p=0.0015$ PTA-adjusted.
\end{table}

\begin{table}[!htbp]
\caption{The fast-mode fraction on phase-randomized surrogates of the activations.}
\label{tab:S-surr}
\centering
\footnotesize
\setlength{\tabcolsep}{3pt}
\begin{tabular}{llcccccc}
\toprule
Feature & Analysis & $r$ D1 & $r$ D2 & $r$ D3 & Pooled $r$ & $p$ & Same sign \\
\midrule
Original fraction            & unadjusted   & 0.27 & 0.41 & 0.44 & 0.31 & $3\times10^{-5}$ & yes \\
                             & PTA-adjusted & 0.23 & 0.32 & 0.32 & 0.26 & 0.002 & yes \\
Surrogate fraction (mean of 5) & unadjusted & 0.28 & 0.34 & 0.38 & 0.30 & $3\times10^{-5}$ & yes \\
                             & PTA-adjusted & 0.22 & 0.26 & 0.28 & 0.23 & 0.007 & yes \\
Original $-$ surrogate       & unadjusted   & $-0.10$ & 0.28 & $-0.01$ & $-0.04$ & 0.92 & no \\
                             & PTA-adjusted & 0.03 & 0.17 & $-0.13$ & 0.03 & 0.79 & no \\
\bottomrule
\end{tabular}
\\[2pt]\raggedright\footnotesize Multivariate phase-randomized surrogates (Methods) preserve the auto- and cross-spectra of the 7 activations and remove higher-order temporal structure; the operator is refit on each surrogate and the fast-mode fraction recomputed (five surrogates per subject, averaged). $r$: rank-biserial group effect (positive = higher in tinnitus); PTA-adjusted: + age, sex where available; pooled $r$: inverse-variance; $p$: Stouffer. Per subject, the surrogate fraction tracks the original closely (Spearman $\rho=$ 0.97 / 0.94 / 0.94 in D1 / D2 / D3; mean absolute difference 0.02 / 0.03 / 0.03; surrogate-to-surrogate SD 0.01). The group difference is fully present in the surrogates, and the residual nonlinear part carries none, so the fast-mode fraction measures the second-order (cross-spectral) structure of the microstate activations.
\end{table}

\begin{figure*}[!htbp]
\centering
% \figfile{figS_channel_spectral_topography}{0.95\textwidth}
\includegraphics[width=0.95\textwidth]{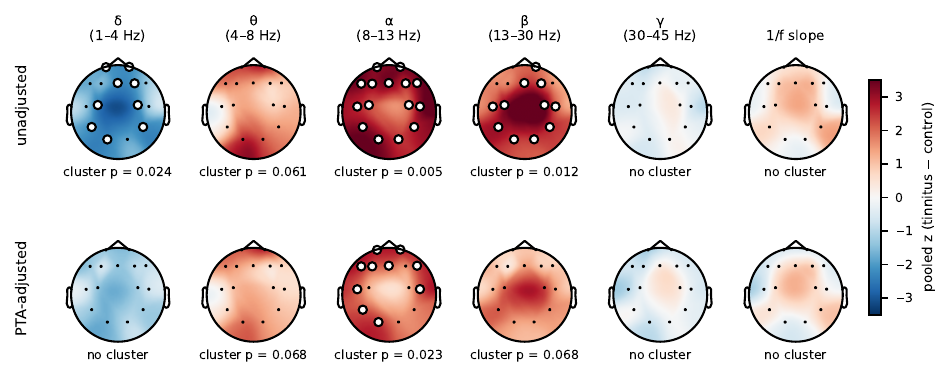}
\caption{Channel-level spectral group differences. Pooled group effect (Stouffer $Z$ over D1--D3) at each of the 15 common channels for each spectral feature, without adjustment (top) and adjusted for PTA (+ age, sex; bottom). Positive: higher in tinnitus. Marked channels: clusters with permutation $p<0.05$ (Methods). After PTA adjustment only relative $\alpha$ forms a significant cluster (10 channels, $p=0.023$), which does not survive correction across the six features ($q=0.14$).}
\label{fig:S-topo}
\end{figure*}

\begin{figure*}[!htbp]
\centering
% \figfile{figS_alpha_cluster}{0.95\textwidth}
\includegraphics[width=0.95\textwidth]{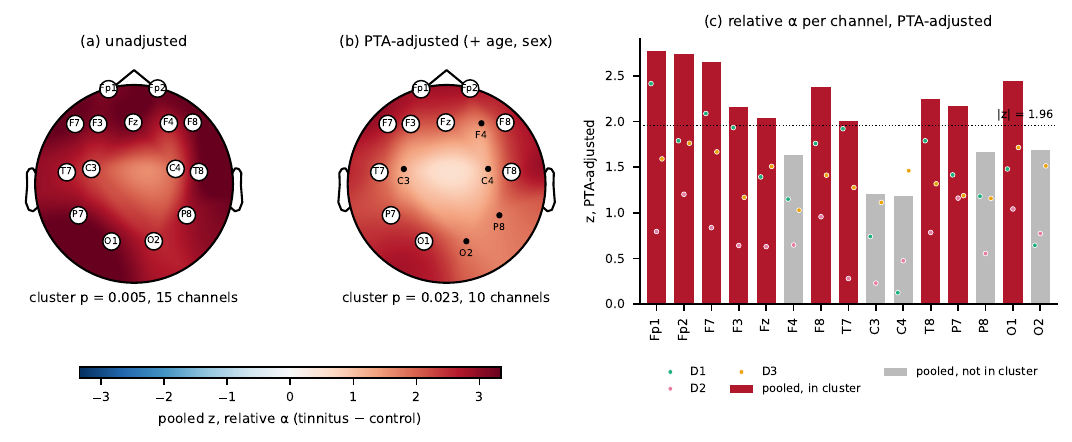}
\caption{The hearing-adjusted relative $\alpha$ cluster. (a), (b) Pooled group effect on relative $\alpha$ power (Stouffer $Z$ over D1--D3) at each of the 15 common channels, without adjustment and adjusted for PTA (+ age, sex); labeled white circles: channels of the significant cluster, black dots: channels outside it. (c) PTA-adjusted effect at each channel: bars, pooled $Z$ (red: in the cluster); points, per-dataset $z$. Dotted line: the cluster-forming threshold $|Z|=1.96$. Positive: higher relative $\alpha$ in tinnitus. The effect is positive at every channel; the channels above threshold lie along the frontal and lateral scalp, but a cluster test does not establish where an effect is located~[citation: Sassenhagen \& Draschkow 2019], and the cluster does not survive correction across the six spectral features ($q=0.14$).}
\label{fig:S-alpha}
\end{figure*}

\begin{table}[!htbp]
\caption{Fast-mode mass and composition by observable type and by microstate.}
\label{tab:S-decomp}
\centering
\scriptsize
\setlength{\tabcolsep}{2.5pt}
\begin{tabular}{lcccccccccccc}
\toprule
 & \multicolumn{6}{c}{Unadjusted} & \multicolumn{6}{c}{PTA-adjusted} \\
\cmidrule(lr){2-7}\cmidrule(lr){8-13}
Feature & $r$ D1 & $r$ D2 & $r$ D3 & $z$ & $q$ & Rep. & $r$ D1 & $r$ D2 & $r$ D3 & $z$ & $q$ & Rep. \\
\midrule
\multicolumn{13}{l}{\emph{fast-mode mass by type}} \\
\quad linear & $-0.28$ & $-0.07$ & $-0.24$ & $-2.95$ & 0.003 & $\checkmark$ & $-0.22$ & $-0.02$ & $-0.29$ & $-2.13$ & 0.033 & $\checkmark$ \\
\quad squared & 0.25 & 0.32 & 0.25 & 3.49 & $7\times10^{-4}$ & $\checkmark$ & 0.20 & 0.24 & 0.23 & 2.42 & 0.023 & $\checkmark$ \\
\quad cross & 0.32 & 0.37 & 0.51 & 4.88 & $3\times10^{-6}$ & $\checkmark$ & 0.27 & 0.32 & 0.34 & 3.41 & 0.002 & $\checkmark$ \\
\multicolumn{13}{l}{\emph{fast-mode mass by map}} \\
\quad map A & 0.15 & 0.50 & 0.45 & 3.94 & $2\times10^{-4}$ & $\checkmark$ & 0.13 & 0.38 & 0.31 & 2.63 & 0.020 & $\checkmark$ \\
\quad map B & 0.31 & 0.35 & 0.46 & 4.63 & $3\times10^{-5}$ & $\checkmark$ & 0.27 & 0.19 & 0.44 & 3.26 & 0.008 & $\checkmark$ \\
\quad map C & 0.21 & 0.45 & 0.52 & 4.36 & $5\times10^{-5}$ & $\checkmark$ & 0.13 & 0.38 & 0.44 & 2.96 & 0.011 & $\checkmark$ \\
\quad map D & 0.21 & 0.29 & 0.25 & 3.10 & 0.003 & $\checkmark$ & 0.15 & 0.26 & 0.14 & 1.92 & 0.055 &  \\
\quad map E & 0.25 & 0.26 & 0.03 & 2.66 & 0.008 & $\checkmark$ & 0.27 & 0.26 & $-0.04$ & 2.17 & 0.053 &  \\
\quad map F & 0.08 & 0.49 & 0.37 & 3.19 & 0.003 & $\checkmark$ & 0.06 & 0.38 & 0.24 & 1.96 & 0.055 &  \\
\quad map G & 0.12 & 0.36 & 0.25 & 2.72 & 0.008 & $\checkmark$ & 0.12 & 0.23 & 0.25 & 1.94 & 0.055 &  \\
\multicolumn{13}{l}{\emph{fast-mode composition by type}} \\
\quad linear & $-0.28$ & $-0.22$ & $-0.35$ & $-3.72$ & $3\times10^{-4}$ & $\checkmark$ & $-0.22$ & $-0.17$ & $-0.33$ & $-2.59$ & 0.014 & $\checkmark$ \\
\quad squared & 0.21 & 0.20 & 0.02 & 2.12 & 0.034 & $\checkmark$ & 0.15 & 0.08 & 0.13 & 1.44 & 0.15 &  \\
\quad cross & 0.29 & 0.24 & 0.50 & 4.33 & $4\times10^{-5}$ & $\checkmark$ & 0.24 & 0.32 & 0.42 & 3.39 & 0.002 & $\checkmark$ \\
\multicolumn{13}{l}{\emph{fast-mode composition by map}} \\
\quad map A & $-0.04$ & 0.29 & 0.33 & 1.56 & 0.53 &  & $-0.02$ & 0.28 & 0.22 & 1.16 & 0.78 &  \\
\quad map B & 0.11 & 0.03 & 0.20 & 1.44 & 0.53 &  & 0.11 & $-0.11$ & 0.28 & 1.07 & 0.78 &  \\
\quad map C & $-0.07$ & 0.05 & 0.33 & 0.62 & 0.85 &  & $-0.12$ & 0.04 & 0.37 & 0.32 & 0.84 &  \\
\quad map D & 0.01 & $-0.25$ & $-0.04$ & $-0.83$ & 0.85 &  & $-0.04$ & $-0.11$ & $-0.09$ & $-0.77$ & 0.78 &  \\
\quad map E & 0.14 & $-0.10$ & $-0.13$ & 0.32 & 0.85 &  & 0.20 & 0.01 & $-0.15$ & 0.84 & 0.78 &  \\
\quad map F & $-0.15$ & 0.15 & 0.28 & 0.21 & 0.85 &  & $-0.14$ & 0.04 & 0.21 & $-0.20$ & 0.84 &  \\
\quad map G & $-0.07$ & 0.11 & $-0.01$ & $-0.18$ & 0.85 &  & $-0.03$ & 0.14 & 0.09 & 0.40 & 0.84 &  \\
\bottomrule
\end{tabular}
\\[2pt]\raggedright\footnotesize $r$: rank-biserial group effect per dataset (positive = higher in tinnitus); $z$: Stouffer over the three datasets; $q$: Benjamini--Hochberg within the feature set; Rep.: replicates (same sign in all three datasets and $q<0.05$). PTA-adjusted: + age and sex where available (Methods, \emph{Pooling and correction}). Mass: a mode's eigenvector energy share on the given observable type (linear $a_k$, squared $a_k^2$, cross $a_ka_l$) or on the observables of a map, summed over the fast modes and divided by the number of retained modes, so that the masses add up to the fast-mode fraction. Composition: the mean share over a subject's fast modes. Notebook Table~21.
\end{table}

\begin{table}[!htbp]
\caption{Microstate and activation-spectral measures: correlation with the fast-mode fraction, and group replication of the spectral measures.}
\label{tab:S-corr}
\centering
\scriptsize
\setlength{\tabcolsep}{2.5pt}
\begin{tabular}{lccccccccc}
\toprule
 & \multicolumn{3}{c}{$\rho$ with fast-mode fraction} & \multicolumn{3}{c}{Group effect} & \multicolumn{3}{c}{Group effect, PTA-adjusted} \\
\cmidrule(lr){2-4}\cmidrule(lr){5-7}\cmidrule(lr){8-10}
Measure & D1 & D2 & D3 & $z$ & $q$ & Rep. & $z$ & $q$ & Rep. \\
\midrule
mean segment duration (ms) & 0.66 & $-0.37$ & $-0.03$ & & &  & & &  \\
segments per s & $-0.66$ & 0.37 & 0.03 & & &  & & &  \\
microstate PC1 & 0.06 & 0.16 & $-0.12$ & & &  & & &  \\
microstate PC2 & 0.44 & 0.27 & $-0.07$ & & &  & & &  \\
microstate PC3 & 0.12 & 0.06 & 0.09 & & &  & & &  \\
rel. $\delta$ (1--4 Hz) & $-0.48$ & $-0.69$ & $-0.17$ & $-2.68$ & 0.015 &  & $-1.89$ & 0.16 &  \\
rel. $\theta$ (4--8 Hz) & $-0.01$ & 0.38 & 0.22 & 1.08 & 0.42 &  & 0.90 & 0.55 &  \\
rel. $\alpha$ (8--13 Hz) & 0.68 & 0.76 & 0.35 & 2.81 & 0.015 & $\checkmark$ & 1.75 & 0.16 &  \\
rel. $\beta$ (13--30 Hz) & $-0.24$ & 0.46 & 0.41 & 3.04 & 0.014 & $\checkmark$ & 1.89 & 0.16 &  \\
rel. $\gamma$ (30--45 Hz) & $-0.80$ & 0.26 & $-0.01$ & $-0.04$ & 0.97 &  & 0.12 & 0.91 &  \\
1/f slope & $-0.69$ & 0.27 & $-0.16$ & 0.83 & 0.49 &  & 0.70 & 0.58 &  \\
strongest microstate statistic (duration A) & 0.58 & -- & -- & & &  & & &  \\
strongest microstate statistic (coverage A) & -- & 0.52 & -- & & &  & & &  \\
strongest microstate statistic ($P(\mathrm{B}\to\mathrm{F})$) & -- & -- & 0.30 & & &  & & &  \\
\bottomrule
\end{tabular}
\\[2pt]\raggedright\footnotesize Left: Spearman $\rho$ over all subjects of a dataset between the fast-mode fraction and the measure. Mean segment duration and segment rate are from the winner-take-all sequence; microstate PCs are the first three principal components of the 56 microstate statistics; relative band power and 1/$f$ slope (2--40~Hz) are from the summed Welch spectrum of the seven activations on the same 60~s window the operator is fit to; the last rows give, per dataset, the microstate statistic with the largest $|\rho|$. Right (spectral measures only): tinnitus--control group effect pooled over the three datasets (Stouffer $z$, Benjamini--Hochberg $q$ over the six measures, Rep.: same sign in all datasets and $q<0.05$), unadjusted and after regressing out PTA (+ age and sex where available); per-dataset effects in notebook Table~22c. Notebook Tables~22a and 22c.
\end{table}

\begin{table}[p]
\caption{Group tests and replication of the 56 microstate statistics, within each dataset and pooled, before and after PTA adjustment.}
\label{tab:S-ms}
\centering
\scriptsize
\setlength{\tabcolsep}{2pt}
\begin{tabular}{lccccccccccccccc}
\toprule
 & \multicolumn{2}{c}{D1} & \multicolumn{2}{c}{D2} & \multicolumn{2}{c}{D3} & \multicolumn{3}{c}{Pooled} & \multicolumn{6}{c}{Pooled, PTA-adjusted} \\
\cmidrule(lr){2-3}\cmidrule(lr){4-5}\cmidrule(lr){6-7}\cmidrule(lr){8-10}\cmidrule(lr){11-16}
Statistic & $\rrb$ & $q$ & $\rrb$ & $q$ & $\rrb$ & $q$ & $z$ & $q$ & Rep. & $r$ D1 & $r$ D2 & $r$ D3 & $z$ & $q$ & Rep. \\
\midrule
\multicolumn{16}{l}{\emph{Coverage}} \\
\quad coverage A & $-0.04$ & 0.84 & 0.19 & 0.67 & 0.32 & 0.64 & 1.24 & 0.57 &  & 0.05 & 0.06 & 0.32 & 1.26 & 0.98 &  \\
\quad coverage B & $-0.14$ & 0.30 & 0.01 & 0.99 & 0.14 & 0.81 & $-0.55$ & 0.86 &  & $-0.10$ & $-0.00$ & 0.18 & $-0.17$ & 0.98 &  \\
\quad coverage C & $-0.13$ & 0.32 & $-0.12$ & 0.71 & $-0.10$ & 0.84 & $-1.55$ & 0.36 &  & $-0.01$ & $-0.13$ & $-0.06$ & $-0.49$ & 0.98 &  \\
\quad coverage D & 0.13 & 0.31 & 0.11 & 0.74 & $-0.19$ & 0.74 & 0.69 & 0.82 &  & 0.05 & 0.04 & $-0.34$ & $-0.47$ & 0.98 &  \\
\quad coverage E & 0.04 & 0.84 & 0.13 & 0.71 & 0.21 & 0.74 & 1.29 & 0.55 &  & 0.02 & 0.18 & 0.13 & 0.91 & 0.98 &  \\
\quad coverage F & $-0.11$ & 0.36 & $-0.25$ & 0.67 & $-0.07$ & 0.86 & $-1.80$ & 0.25 &  & $-0.06$ & $-0.21$ & 0.06 & $-0.72$ & 0.98 &  \\
\quad coverage G & 0.22 & 0.14 & 0.27 & 0.65 & 0.13 & 0.81 & 2.75 & 0.063 &  & 0.09 & 0.28 & $-0.00$ & 1.24 & 0.98 &  \\
\multicolumn{16}{l}{\emph{Mean duration}} \\
\quad duration A & 0.08 & 0.58 & $-0.30$ & 0.64 & 0.32 & 0.64 & 0.60 & 0.83 &  & 0.08 & $-0.26$ & 0.33 & 0.70 & 0.98 &  \\
\quad duration B & $-0.01$ & 0.99 & $-0.27$ & 0.65 & 0.24 & 0.74 & $-0.22$ & 0.89 &  & 0.03 & $-0.22$ & 0.30 & 0.40 & 0.98 &  \\
\quad duration C & 0.00 & 0.99 & $-0.44$ & 0.18 & 0.09 & 0.85 & $-1.07$ & 0.66 &  & 0.04 & $-0.33$ & 0.16 & $-0.19$ & 0.98 &  \\
\quad duration D & 0.15 & 0.30 & $-0.29$ & 0.64 & 0.02 & 0.97 & 0.26 & 0.89 &  & 0.09 & $-0.25$ & $-0.05$ & $-0.19$ & 0.98 &  \\
\quad duration E & 0.06 & 0.71 & $-0.20$ & 0.67 & 0.20 & 0.74 & 0.43 & 0.87 &  & 0.03 & $-0.18$ & 0.16 & 0.15 & 0.98 &  \\
\quad duration F & 0.03 & 0.87 & $-0.40$ & 0.27 & 0.06 & 0.86 & $-0.85$ & 0.76 &  & 0.02 & $-0.29$ & 0.13 & $-0.24$ & 0.98 &  \\
\quad duration G & 0.14 & 0.30 & $-0.17$ & 0.67 & 0.21 & 0.74 & 1.09 & 0.66 &  & 0.06 & $-0.14$ & 0.15 & 0.44 & 0.98 &  \\
\multicolumn{16}{l}{\emph{Transition probability}} \\
\quad $P(\mathrm{A}\to\mathrm{B})$ & $-0.19$ & 0.14 & 0.18 & 0.67 & 0.15 & 0.81 & $-0.41$ & 0.87 &  & $-0.14$ & 0.09 & 0.37 & 0.34 & 0.98 &  \\
\quad $P(\mathrm{A}\to\mathrm{C})$ & $-0.07$ & 0.69 & $-0.23$ & 0.67 & 0.35 & 0.64 & $-0.17$ & 0.89 &  & $-0.01$ & $-0.20$ & 0.40 & 0.49 & 0.98 &  \\
\quad $P(\mathrm{A}\to\mathrm{D})$ & 0.21 & 0.14 & 0.54 & 0.12 & 0.21 & 0.74 & 3.77 & 0.009 & $\checkmark$ & 0.18 & 0.45 & 0.07 & 2.41 & 0.88 &  \\
\quad $P(\mathrm{A}\to\mathrm{E})$ & $-0.04$ & 0.83 & $-0.03$ & 0.96 & 0.40 & 0.64 & 0.77 & 0.80 &  & $-0.02$ & 0.04 & 0.39 & 1.01 & 0.98 &  \\
\quad $P(\mathrm{A}\to\mathrm{F})$ & $-0.05$ & 0.80 & $-0.02$ & 0.98 & 0.27 & 0.71 & 0.35 & 0.89 &  & 0.06 & 0.19 & 0.33 & 1.67 & 0.98 &  \\
\quad $P(\mathrm{A}\to\mathrm{G})$ & 0.20 & 0.14 & 0.22 & 0.67 & 0.20 & 0.74 & 2.66 & 0.063 &  & 0.17 & $-0.02$ & 0.13 & 1.32 & 0.98 &  \\
\quad $P(\mathrm{B}\to\mathrm{A})$ & $-0.10$ & 0.47 & 0.15 & 0.69 & 0.15 & 0.81 & 0.21 & 0.89 &  & $-0.05$ & 0.03 & 0.33 & 0.60 & 0.98 &  \\
\quad $P(\mathrm{B}\to\mathrm{C})$ & $-0.12$ & 0.33 & 0.01 & 0.99 & $-0.02$ & 0.97 & $-0.89$ & 0.75 &  & $-0.10$ & 0.13 & 0.07 & $-0.08$ & 0.98 &  \\
\quad $P(\mathrm{B}\to\mathrm{D})$ & 0.10 & 0.40 & 0.06 & 0.87 & $-0.08$ & 0.85 & 0.67 & 0.82 &  & 0.01 & 0.10 & $-0.17$ & $-0.11$ & 0.98 &  \\
\quad $P(\mathrm{B}\to\mathrm{E})$ & $-0.11$ & 0.38 & 0.12 & 0.71 & 0.22 & 0.74 & 0.20 & 0.89 &  & $-0.06$ & 0.04 & 0.23 & 0.34 & 0.98 &  \\
\quad $P(\mathrm{B}\to\mathrm{F})$ & $-0.14$ & 0.30 & 0.06 & 0.85 & 0.00 & 1.0 & $-0.83$ & 0.76 &  & $-0.07$ & 0.10 & 0.06 & $-0.02$ & 0.98 &  \\
\quad $P(\mathrm{B}\to\mathrm{G})$ & 0.05 & 0.80 & 0.13 & 0.71 & 0.34 & 0.64 & 1.75 & 0.26 &  & $-0.01$ & 0.24 & 0.25 & 1.20 & 0.98 &  \\
\quad $P(\mathrm{C}\to\mathrm{A})$ & $-0.11$ & 0.36 & $-0.09$ & 0.79 & 0.28 & 0.65 & $-0.26$ & 0.89 &  & $-0.07$ & $-0.09$ & 0.38 & 0.35 & 0.98 &  \\
\quad $P(\mathrm{C}\to\mathrm{B})$ & $-0.05$ & 0.81 & $-0.08$ & 0.83 & $-0.04$ & 0.93 & $-0.69$ & 0.82 &  & $-0.00$ & 0.07 & 0.07 & 0.31 & 0.98 &  \\
\quad $P(\mathrm{C}\to\mathrm{D})$ & 0.09 & 0.52 & $-0.06$ & 0.85 & $-0.12$ & 0.81 & 0.10 & 0.92 &  & 0.07 & $-0.25$ & $-0.11$ & $-0.43$ & 0.98 &  \\
\quad $P(\mathrm{C}\to\mathrm{E})$ & $-0.08$ & 0.62 & 0.26 & 0.67 & 0.14 & 0.81 & 0.63 & 0.82 &  & $-0.02$ & 0.25 & 0.13 & 0.82 & 0.98 &  \\
\quad $P(\mathrm{C}\to\mathrm{F})$ & $-0.12$ & 0.33 & $-0.17$ & 0.67 & $-0.17$ & 0.81 & $-1.88$ & 0.24 &  & $-0.05$ & $-0.01$ & $-0.07$ & $-0.52$ & 0.98 &  \\
\quad $P(\mathrm{C}\to\mathrm{G})$ & $-0.03$ & 0.88 & 0.21 & 0.67 & $-0.10$ & 0.84 & 0.15 & 0.90 &  & $-0.04$ & 0.11 & 0.01 & 0.05 & 0.98 &  \\
\quad $P(\mathrm{D}\to\mathrm{A})$ & 0.13 & 0.31 & 0.54 & 0.12 & 0.20 & 0.74 & 3.18 & 0.041 & $\checkmark$ & 0.10 & 0.46 & 0.06 & 1.96 & 0.94 &  \\
\quad $P(\mathrm{D}\to\mathrm{B})$ & 0.00 & 0.99 & 0.17 & 0.67 & $-0.12$ & 0.81 & 0.18 & 0.89 &  & $-0.07$ & 0.33 & $-0.18$ & $-0.07$ & 0.98 &  \\
\quad $P(\mathrm{D}\to\mathrm{C})$ & 0.03 & 0.87 & 0.17 & 0.67 & $-0.15$ & 0.81 & 0.31 & 0.89 &  & 0.02 & 0.01 & $-0.13$ & $-0.23$ & 0.98 &  \\
\quad $P(\mathrm{D}\to\mathrm{E})$ & 0.26 & 0.067 & 0.17 & 0.67 & 0.12 & 0.81 & 2.72 & 0.063 &  & 0.15 & 0.22 & $-0.01$ & 1.47 & 0.98 &  \\
\quad $P(\mathrm{D}\to\mathrm{F})$ & 0.03 & 0.87 & $-0.04$ & 0.92 & $-0.37$ & 0.64 & $-0.97$ & 0.71 &  & 0.01 & $-0.18$ & $-0.38$ & $-1.38$ & 0.98 &  \\
\quad $P(\mathrm{D}\to\mathrm{G})$ & 0.19 & 0.14 & 0.16 & 0.69 & 0.10 & 0.84 & 2.12 & 0.17 &  & 0.07 & 0.12 & $-0.05$ & 0.63 & 0.98 &  \\
\quad $P(\mathrm{E}\to\mathrm{A})$ & $-0.01$ & 0.99 & $-0.17$ & 0.67 & 0.34 & 0.64 & 0.43 & 0.87 &  & 0.04 & $-0.16$ & 0.33 & 0.67 & 0.98 &  \\
\quad $P(\mathrm{E}\to\mathrm{B})$ & $-0.15$ & 0.30 & 0.24 & 0.67 & 0.19 & 0.74 & 0.21 & 0.89 &  & $-0.11$ & 0.19 & 0.20 & 0.28 & 0.98 &  \\
\quad $P(\mathrm{E}\to\mathrm{C})$ & $-0.04$ & 0.83 & 0.24 & 0.67 & 0.08 & 0.85 & 0.64 & 0.82 &  & 0.01 & 0.14 & 0.16 & 0.82 & 0.98 &  \\
\quad $P(\mathrm{E}\to\mathrm{D})$ & 0.21 & 0.14 & 0.13 & 0.71 & 0.12 & 0.81 & 2.22 & 0.15 &  & 0.08 & 0.18 & $-0.03$ & 0.84 & 0.98 &  \\
\quad $P(\mathrm{E}\to\mathrm{F})$ & $-0.01$ & 0.99 & $-0.15$ & 0.70 & 0.04 & 0.93 & $-0.42$ & 0.87 &  & $-0.01$ & $-0.14$ & 0.10 & $-0.13$ & 0.98 &  \\
\quad $P(\mathrm{E}\to\mathrm{G})$ & 0.14 & 0.30 & 0.20 & 0.67 & 0.29 & 0.64 & 2.49 & 0.091 &  & 0.07 & 0.29 & 0.19 & 1.63 & 0.98 &  \\
\quad $P(\mathrm{F}\to\mathrm{A})$ & 0.04 & 0.84 & $-0.07$ & 0.84 & 0.30 & 0.64 & 0.93 & 0.73 &  & 0.11 & $-0.02$ & 0.37 & 1.54 & 0.98 &  \\
\quad $P(\mathrm{F}\to\mathrm{B})$ & $-0.14$ & 0.30 & $-0.12$ & 0.71 & 0.07 & 0.86 & $-1.17$ & 0.61 &  & $-0.08$ & $-0.11$ & 0.17 & $-0.29$ & 0.98 &  \\
\quad $P(\mathrm{F}\to\mathrm{C})$ & $-0.01$ & 0.99 & $-0.47$ & 0.17 & $-0.18$ & 0.81 & $-2.03$ & 0.20 &  & 0.07 & $-0.39$ & $-0.08$ & $-0.77$ & 0.98 &  \\
\quad $P(\mathrm{F}\to\mathrm{D})$ & $-0.00$ & 0.99 & 0.13 & 0.71 & $-0.30$ & 0.64 & $-0.49$ & 0.87 &  & $-0.04$ & 0.10 & $-0.31$ & $-0.77$ & 0.98 &  \\
\quad $P(\mathrm{F}\to\mathrm{E})$ & $-0.05$ & 0.82 & $-0.12$ & 0.71 & 0.06 & 0.86 & $-0.52$ & 0.87 &  & $-0.02$ & $-0.12$ & 0.10 & $-0.17$ & 0.98 &  \\
\quad $P(\mathrm{F}\to\mathrm{G})$ & 0.01 & 0.99 & 0.52 & 0.12 & 0.08 & 0.85 & 1.84 & 0.24 &  & $-0.02$ & 0.40 & 0.02 & 0.94 & 0.98 &  \\
\quad $P(\mathrm{G}\to\mathrm{A})$ & 0.10 & 0.40 & 0.45 & 0.18 & 0.21 & 0.74 & 2.71 & 0.063 &  & 0.13 & 0.35 & 0.16 & 2.08 & 0.94 &  \\
\quad $P(\mathrm{G}\to\mathrm{B})$ & 0.15 & 0.30 & $-0.00$ & 0.99 & 0.30 & 0.64 & 1.96 & 0.22 &  & 0.05 & $-0.14$ & 0.21 & 0.48 & 0.98 &  \\
\quad $P(\mathrm{G}\to\mathrm{C})$ & 0.00 & 0.99 & 0.33 & 0.54 & $-0.01$ & 1.0 & 1.04 & 0.67 &  & 0.02 & 0.28 & 0.05 & 0.95 & 0.98 &  \\
\quad $P(\mathrm{G}\to\mathrm{D})$ & 0.14 & 0.30 & 0.30 & 0.64 & 0.13 & 0.81 & 2.27 & 0.14 &  & 0.03 & 0.37 & $-0.00$ & 1.10 & 0.98 &  \\
\quad $P(\mathrm{G}\to\mathrm{E})$ & 0.17 & 0.26 & 0.18 & 0.67 & 0.31 & 0.64 & 2.67 & 0.063 &  & 0.07 & 0.26 & 0.21 & 1.61 & 0.98 &  \\
\quad $P(\mathrm{G}\to\mathrm{F})$ & 0.07 & 0.68 & 0.22 & 0.67 & 0.15 & 0.81 & 1.58 & 0.36 &  & 0.04 & 0.03 & 0.09 & 0.52 & 0.98 &  \\
\bottomrule
\end{tabular}
\\[2pt]\raggedright\footnotesize D1--D3: rank-biserial effect $\rrb$ of the two-sided Mann--Whitney test (positive = higher in tinnitus) and Benjamini--Hochberg $q$ within the dataset over the 56 statistics; no statistic reached $q<0.05$ in any dataset. Pooled: Stouffer $z$ over the three datasets, $q$ over the 56 statistics, and Rep.: replicates (same sign in all three datasets and $q<0.05$). PTA-adjusted: partial effect $r$ after regressing out PTA (+ age and sex where available; Methods, \emph{Pooling and correction}), then pooled the same way. Group medians are in notebook Table~3. Notebook Tables~3 and 18b.
\end{table}

\begin{table}[!htbp]
\caption{Global explained variance of the template maps on each dataset.}
\label{tab:S-gev}
\centering
\footnotesize
\setlength{\tabcolsep}{3pt}
\begin{tabular}{llcccccccc}
\toprule
Dataset & Maps & A & B & C & D & E & F & G & Total \\
\midrule
D1 & template on own montage & 0.087 & 0.088 & 0.116 & 0.086 & 0.071 & 0.057 & 0.058 & 0.563 \\
 & template via D2 & 0.087 & 0.089 & 0.111 & 0.083 & 0.073 & 0.058 & 0.058 & 0.559 \\
 & template via D3 & 0.087 & 0.089 & 0.116 & 0.083 & 0.072 & 0.058 & 0.058 & 0.563 \\
D2 & template on own montage & 0.071 & 0.076 & 0.108 & 0.092 & 0.055 & 0.075 & 0.054 & 0.531 \\
 & template via D1 & 0.071 & 0.076 & 0.109 & 0.092 & 0.055 & 0.075 & 0.054 & 0.531 \\
 & template via D3 & 0.070 & 0.077 & 0.108 & 0.092 & 0.055 & 0.076 & 0.054 & 0.532 \\
D3 & template on own montage & 0.016 & 0.022 & 0.057 & 0.035 & 0.015 & 0.243 & 0.018 & 0.406 \\
 & template via D1 & 0.016 & 0.021 & 0.058 & 0.035 & 0.015 & 0.243 & 0.018 & 0.406 \\
 & template via D2 & 0.017 & 0.020 & 0.069 & 0.035 & 0.015 & 0.234 & 0.018 & 0.407 \\
\bottomrule
\end{tabular}
\\[2pt]\raggedright\footnotesize GEV of the winner-take-all back-fit, pooled over the subjects of a dataset, per map and in total. ``Template via D$n$'': the template interpolated onto dataset D$n$'s montage and from there onto the test montage, as in the cross-dataset analyses; the route makes no difference. Notebook Table~6.
\end{table}

\begin{table}[!htbp]
\caption{Cross-validated prediction of PTA from the EEG features.}
\label{tab:S-pta}
\centering
\footnotesize
\setlength{\tabcolsep}{3pt}
\begin{tabular}{llcclccc}
\toprule
Dataset & Group & $n$ & PTA SD (dB) & Features & $\rho$(pred., PTA) & CV $R^2$ & $p$ \\
\midrule
D1 & tinnitus & 181 & 15.0 & Microstate (56) & 0.19 & $-0.03$ & 0.005 \\
 & & &  & Koopman $|\lam|$ quantiles & 0.01 & $-0.06$ & 0.13 \\
D1 & control & 72 & 4.8 & Microstate (56) & 0.28 & $-0.42$ & 0.025 \\
 & & &  & Koopman $|\lam|$ quantiles & 0.21 & $-0.33$ & 0.015 \\
D1 & all & 253 & 15.2 & Microstate (56) & 0.24 & 0.01 & 0.005 \\
 & & &  & Koopman $|\lam|$ quantiles & 0.17 & $-0.12$ & 0.005 \\
D2 & tinnitus & 31 & 12.6 & Microstate (56) & $-0.39$ & $-1.41$ & 0.81 \\
 & & &  & Koopman $|\lam|$ quantiles & $-0.33$ & $-0.11$ & 0.68 \\
D2 & all & 43 & 13.6 & Microstate (56) & $-0.30$ & $-0.03$ & 0.66 \\
 & & &  & Koopman $|\lam|$ quantiles & $-0.11$ & $-0.02$ & 0.28 \\
D3 & tinnitus & 22 & 16.1 & Microstate (56) & $-0.27$ & $-0.20$ & 0.45 \\
 & & &  & Koopman $|\lam|$ quantiles & $-0.35$ & $-0.11$ & 0.66 \\
D3 & control & 15 & 5.3 & Microstate (56) & $-0.50$ & $-0.44$ & 0.70 \\
 & & &  & Koopman $|\lam|$ quantiles & $-0.45$ & $-0.45$ & 0.64 \\
D3 & all & 37 & 14.6 & Microstate (56) & $-0.09$ & $-0.04$ & 0.26 \\
 & & &  & Koopman $|\lam|$ quantiles & 0.10 & $-0.03$ & 0.070 \\
\bottomrule
\end{tabular}
\\[2pt]\raggedright\footnotesize Ridge regression (5-fold cross-validation, $\alpha$ chosen by inner cross-validation) of PTA on the 56 microstate statistics or on the 21 Koopman $|\lam|$ quantile features; $\rho$: Spearman correlation of the out-of-fold prediction with PTA; $p$: permutation test (200 permutations of PTA). Groups with fewer than 15 subjects with PTA are omitted. Notebook Table~12.
\end{table}

\begin{table}[!htbp]
\caption{Correlation of the classifier scores with tinnitus severity.}
\label{tab:S-sev}
\centering
\footnotesize
\setlength{\tabcolsep}{3pt}
\begin{tabular}{llllll}
\toprule
Dataset & Scale & $n$ & Scores & Microstate SVM & Koopman \\
\midrule
D1 & THI & 180 & trained on D2 & $-0.01$ (0.85) & $-0.01$ (0.90) \\
 & &  & trained on D3 & 0.09 (0.23) & 0.02 (0.80) \\
 & &  & within-dataset CV & $-0.13$ (0.078) & 0.03 (0.67) \\
D2 & THI grade (S2) & 29 & trained on D1 & $-0.28$ (0.14) & 0.11 (0.58) \\
 & &  & trained on D3 & 0.05 (0.81) & 0.07 (0.71) \\
 & &  & within-dataset CV & 0.30 (0.11) & $-0.05$ (0.81) \\
D3 & TFI & 22 & trained on D1 & $-0.06$ (0.79) & $-0.09$ (0.70) \\
 & &  & trained on D2 & 0.06 (0.80) & $-0.13$ (0.57) \\
 & &  & within-dataset CV & 0.48 (0.025) & $-0.16$ (0.49) \\
\bottomrule
\end{tabular}
\\[2pt]\raggedright\footnotesize Spearman $\rho$ ($p$) between the classifier score and the severity score, tinnitus subjects only. THI: Tinnitus Handicap Inventory (D2: THI grade); TFI: Tinnitus Functional Index. No correlation survives Benjamini--Hochberg correction over the 18 tests (all $q\geq0.46$). Notebook Table~8.
\end{table}

\begin{table}[p]
\caption{Per-feature replication of the Koopman and EEG-spectral features, before and after PTA adjustment.}
\label{tab:S-perfeat-koop}
\centering
\scriptsize
\setlength{\tabcolsep}{2.5pt}
\begin{tabular}{lcccccccccccc}
\toprule
 & \multicolumn{6}{c}{Unadjusted} & \multicolumn{6}{c}{PTA-adjusted} \\
\cmidrule(lr){2-7}\cmidrule(lr){8-13}
Feature & $r$ D1 & $r$ D2 & $r$ D3 & $z$ & $q$ & Rep. & $r$ D1 & $r$ D2 & $r$ D3 & $z$ & $q$ & Rep. \\
\midrule
\multicolumn{13}{l}{\emph{Koopman $|\lam|$ quantiles (21)}} \\
\quad $|\lam|$ q0.00 & $-0.09$ & $-0.22$ & $-0.32$ & $-2.22$ & 0.030 & $\checkmark$ & $-0.12$ & $-0.10$ & $-0.36$ & $-1.90$ & 0.067 &  \\
\quad $|\lam|$ q0.05 & $-0.00$ & $-0.35$ & $-0.39$ & $-2.21$ & 0.030 & $\checkmark$ & 0.00 & $-0.22$ & $-0.42$ & $-1.62$ & 0.12 &  \\
\quad $|\lam|$ q0.11 & $-0.03$ & $-0.45$ & $-0.47$ & $-2.96$ & 0.004 & $\checkmark$ & $-0.06$ & $-0.28$ & $-0.42$ & $-2.17$ & 0.037 & $\checkmark$ \\
\quad $|\lam|$ q0.16 & $-0.05$ & $-0.50$ & $-0.56$ & $-3.56$ & $5\times10^{-4}$ & $\checkmark$ & $-0.07$ & $-0.36$ & $-0.52$ & $-2.79$ & 0.007 & $\checkmark$ \\
\quad $|\lam|$ q0.21 & $-0.12$ & $-0.55$ & $-0.53$ & $-4.12$ & $6\times10^{-5}$ & $\checkmark$ & $-0.14$ & $-0.36$ & $-0.55$ & $-3.28$ & 0.003 & $\checkmark$ \\
\quad $|\lam|$ q0.26 & $-0.18$ & $-0.57$ & $-0.55$ & $-4.68$ & $6\times10^{-6}$ & $\checkmark$ & $-0.18$ & $-0.37$ & $-0.53$ & $-3.52$ & 0.003 & $\checkmark$ \\
\quad $|\lam|$ q0.32 & $-0.21$ & $-0.60$ & $-0.50$ & $-4.78$ & $4\times10^{-6}$ & $\checkmark$ & $-0.16$ & $-0.39$ & $-0.49$ & $-3.35$ & 0.003 & $\checkmark$ \\
\quad $|\lam|$ q0.37 & $-0.25$ & $-0.59$ & $-0.52$ & $-5.12$ & $3\times10^{-6}$ & $\checkmark$ & $-0.19$ & $-0.39$ & $-0.48$ & $-3.54$ & 0.003 & $\checkmark$ \\
\quad $|\lam|$ q0.42 & $-0.24$ & $-0.56$ & $-0.52$ & $-4.98$ & $3\times10^{-6}$ & $\checkmark$ & $-0.18$ & $-0.35$ & $-0.47$ & $-3.34$ & 0.003 & $\checkmark$ \\
\quad $|\lam|$ q0.47 & $-0.28$ & $-0.53$ & $-0.45$ & $-4.93$ & $3\times10^{-6}$ & $\checkmark$ & $-0.20$ & $-0.33$ & $-0.36$ & $-3.06$ & 0.005 & $\checkmark$ \\
\quad $|\lam|$ q0.53 & $-0.31$ & $-0.52$ & $-0.35$ & $-4.80$ & $4\times10^{-6}$ & $\checkmark$ & $-0.21$ & $-0.33$ & $-0.26$ & $-2.82$ & 0.007 & $\checkmark$ \\
\quad $|\lam|$ q0.58 & $-0.33$ & $-0.51$ & $-0.38$ & $-5.06$ & $3\times10^{-6}$ & $\checkmark$ & $-0.24$ & $-0.32$ & $-0.25$ & $-2.96$ & 0.005 & $\checkmark$ \\
\quad $|\lam|$ q0.63 & $-0.34$ & $-0.50$ & $-0.41$ & $-5.14$ & $3\times10^{-6}$ & $\checkmark$ & $-0.26$ & $-0.34$ & $-0.32$ & $-3.38$ & 0.003 & $\checkmark$ \\
\quad $|\lam|$ q0.68 & $-0.32$ & $-0.46$ & $-0.42$ & $-4.89$ & $3\times10^{-6}$ & $\checkmark$ & $-0.23$ & $-0.29$ & $-0.34$ & $-3.06$ & 0.005 & $\checkmark$ \\
\quad $|\lam|$ q0.74 & $-0.27$ & $-0.44$ & $-0.50$ & $-4.77$ & $4\times10^{-6}$ & $\checkmark$ & $-0.23$ & $-0.33$ & $-0.38$ & $-3.25$ & 0.003 & $\checkmark$ \\
\quad $|\lam|$ q0.79 & $-0.25$ & $-0.36$ & $-0.51$ & $-4.42$ & $2\times10^{-5}$ & $\checkmark$ & $-0.20$ & $-0.31$ & $-0.44$ & $-3.18$ & 0.004 & $\checkmark$ \\
\quad $|\lam|$ q0.84 & $-0.31$ & $-0.32$ & $-0.36$ & $-4.20$ & $4\times10^{-5}$ & $\checkmark$ & $-0.25$ & $-0.25$ & $-0.30$ & $-2.93$ & 0.005 & $\checkmark$ \\
\quad $|\lam|$ q0.89 & $-0.27$ & $-0.31$ & $-0.41$ & $-4.09$ & $6\times10^{-5}$ & $\checkmark$ & $-0.22$ & $-0.27$ & $-0.39$ & $-3.03$ & 0.005 & $\checkmark$ \\
\quad $|\lam|$ q0.95 & $-0.27$ & $-0.30$ & $-0.47$ & $-4.21$ & $4\times10^{-5}$ & $\checkmark$ & $-0.22$ & $-0.22$ & $-0.39$ & $-2.93$ & 0.005 & $\checkmark$ \\
\quad $|\lam|$ q1.00 & 0.01 & $-0.31$ & $-0.41$ & $-2.05$ & 0.043 &  & $-0.04$ & $-0.12$ & $-0.37$ & $-1.48$ & 0.15 &  \\
\quad fraction in range & 0.28 & $-0.06$ & $-0.13$ & 1.47 & 0.14 &  & 0.24 & 0.03 & $-0.14$ & 1.17 & 0.24 &  \\
\multicolumn{13}{l}{\emph{Koopman $|\lam|$ $\times$ freq cells (70)}} \\
\quad <1 Hz | 0.70--0.90 & $-0.24$ & 0.43 & 0.19 & 0.15 & 0.93 &  & $-0.14$ & 0.25 & 0.06 & $-0.03$ & 0.97 &  \\
\quad <1 Hz | 0.90--0.95 & $-0.23$ & 0.13 & $-0.25$ & $-1.99$ & 0.13 &  & $-0.19$ & 0.18 & $-0.21$ & $-1.26$ & 0.51 &  \\
\quad <1 Hz | 0.95--0.98 & $-0.24$ & $-0.14$ & $-0.07$ & $-2.33$ & 0.075 &  & $-0.16$ & $-0.21$ & $-0.10$ & $-1.72$ & 0.32 &  \\
\quad <1 Hz | 0.98--0.99 & $-0.01$ & $-0.25$ & $-0.10$ & $-1.26$ & 0.39 &  & $-0.01$ & $-0.13$ & $-0.10$ & $-0.66$ & 0.85 &  \\
\quad 1--4 Hz | 0.70--0.90 & $-0.18$ & 0.17 & 0.30 & 0.09 & 0.93 &  & $-0.17$ & 0.07 & 0.23 & $-0.27$ & 0.97 &  \\
\quad 1--4 Hz | 0.90--0.95 & 0.04 & $-0.29$ & 0.09 & $-0.28$ & 0.93 &  & 0.05 & $-0.25$ & 0.17 & 0.18 & 0.97 &  \\
\quad 4--8 Hz | 0.70--0.90 & 0.11 & 0.03 & $-0.11$ & 0.71 & 0.80 &  & 0.08 & $-0.06$ & $-0.05$ & 0.25 & 0.97 &  \\
\quad 8--13 Hz | 0.70--0.90 & 0.04 & $-0.20$ & 0.02 & 0.17 & 0.93 &  & 0.06 & $-0.09$ & 0.10 & 1.00 & 0.59 &  \\
\quad 13--30 Hz | 0.70--0.90 & 0.29 & 0.41 & 0.38 & 4.42 & $1\times10^{-4}$ & $\checkmark$ & 0.26 & 0.27 & 0.40 & 3.30 & 0.015 & $\checkmark$ \\
\quad 13--30 Hz | 0.90--0.95 & 0.18 & 0.06 & 0.12 & 1.95 & 0.13 &  & 0.10 & 0.05 & 0.13 & 1.18 & 0.51 &  \\
\quad 13--30 Hz | 0.95--0.98 & 0.01 & $-0.04$ & 0.12 & 0.46 & 0.88 &  & $-0.02$ & $-0.02$ & 0.13 & $-0.15$ & 0.97 &  \\
\quad 30--50 Hz | 0.70--0.90 & 0.06 & $-0.14$ & $-0.20$ & $-0.57$ & 0.86 &  & 0.08 & 0.07 & $-0.24$ & 0.06 & 0.97 &  \\
\quad 30--50 Hz | 0.90--0.95 & $-0.34$ & $-0.30$ & 0.04 & $-3.28$ & 0.005 &  & $-0.31$ & $-0.26$ & 0.23 & $-1.98$ & 0.24 &  \\
\quad 30--50 Hz | 0.95--0.98 & $-0.05$ & $-0.15$ & $-0.21$ & $-1.75$ & 0.17 &  & 0.02 & $-0.23$ & $-0.39$ & $-1.47$ & 0.42 &  \\
\quad >50 Hz | 0.70--0.90 & 0.27 & 0.51 & 0.19 & 4.06 & $4\times10^{-4}$ & $\checkmark$ & 0.18 & 0.31 & 0.24 & 2.54 & 0.084 &  \\
\multicolumn{13}{l}{\emph{Spectral, channel mean (6)}} \\
\quad rel. $\delta$ (1--4 Hz) & $-0.25$ & $-0.23$ & 0.05 & $-2.36$ & 0.037 &  & $-0.19$ & $-0.21$ & 0.05 & $-1.55$ & 0.25 &  \\
\quad rel. $\theta$ (4--8 Hz) & 0.07 & 0.10 & 0.30 & 1.67 & 0.14 &  & 0.04 & 0.14 & 0.31 & 1.39 & 0.25 &  \\
\quad rel. $\alpha$ (8--13 Hz) & 0.19 & 0.18 & 0.36 & 2.93 & 0.020 & $\checkmark$ & 0.15 & 0.18 & 0.28 & 2.04 & 0.25 &  \\
\quad rel. $\beta$ (13--30 Hz) & 0.11 & 0.35 & 0.19 & 2.43 & 0.037 & $\checkmark$ & 0.08 & 0.22 & 0.16 & 1.44 & 0.25 &  \\
\quad rel. $\gamma$ (30--45 Hz) & $-0.21$ & 0.47 & $-0.16$ & $-0.54$ & 0.71 &  & $-0.17$ & 0.41 & $-0.20$ & $-0.50$ & 0.74 &  \\
\quad 1/f slope & $-0.15$ & 0.58 & $-0.13$ & 0.31 & 0.75 &  & $-0.14$ & 0.46 & $-0.15$ & $-0.04$ & 0.97 &  \\
\bottomrule
\end{tabular}
\\[2pt]\raggedright\footnotesize $r$: rank-biserial group effect per dataset (positive = higher in tinnitus); $z$: Stouffer over the three datasets; $q$: Benjamini--Hochberg within the feature set; Rep.: replicates (same sign in all three datasets and $q<0.05$). PTA-adjusted: + age and sex where available (Methods, \emph{Pooling and correction}). Koopman $|\lam|$ quantiles: 20 quantiles of the retained $|\lam|$ and the fraction retained; $|\lam|\times$frequency cells: fraction of retained modes per cell, cells with fewer than 5 subjects having any mode omitted; spectral: channel-mean relative band power and 1/$f$ slope of the EEG. Notebook Table~18b.
\end{table}

\end{document}